\documentstyle[12pt,aps,epsfig]{revtex}
\begin{document}

\title{Pair Fluctuations in Ultra-small Fermi Systems within Self-Consistent RPA
at Finite Temperature}

\author {
A. Storozhenko $^a$, P. Schuck $^b$, J. Dukelsky $^c$, G.
R\"{o}pke $^d$, A. Vdovin $^a$}


\bigskip
\address{
$^a$ Bogoliubov Laboratory of Theoretical
Physics, Joint Institute for Nuclear Research,\\ 141980 Dubna, Russia \\
$^b$ Institut de Physique Nucleaire,
F-91406 Orsay Cedex, France \\
and \\
Laboratoire de Physique et Mod\'elisation des Milieux Condens\'es,
CNRS $\&$ Universit\'e
Joseph Fourier, Maison des Magist\`eres, B.P. 166, 38042 Grenoble Cedex 9, France\\
$^c$ Instituto de Estructura de la Materia, Serrano 123, 28006
Madrid, Spain \\
$^d$ Fachbereich Physik, Universit\"{a}t Rostock,
Universit\"{a}tsplatz 1, 18051 Rostock, Germany }

\maketitle

\bigskip
\begin{abstract}
A self-consistent version of the Thermal Random Phase
Approximation (TSCRPA) is developed within the Matsubara Green's
Function (GF) formalism. The TSCRPA is applied to the many level
pairing model. The normal phase of the system is considered. The
TSCRPA results are compared with the exact ones calculated for the
Grand Canonical Ensemble. Advantages of the TSCRPA over the
Thermal Mean Field Approximation (TMFA) and the standard Thermal
Random Phase Approximation (TRPA) are demonstrated. Results for
correlation functions, excitation energies, single particle level
densities, etc., as a function of temperature are presented.
\end{abstract}

\section{Introduction}

Pairing properties of finite Fermi systems such as ultrasmall
metallic grains have recently received a great deal of attention.
This has been spurred by a series of spectacular experiments of
Ralph, Black and Tinkham \cite{RBT}. In order to correctly
describe pairing properties it has been recognized that the
finiteness of the systems (grains) needs to consider quantum
fluctuations, good particle number, number parity, etc. seriously,
since the coherence length may be of the order of the system size.
The situation for metallic grains has in the meanwhile been well
described in several review articles \cite{Del1,Del2} (see also
\cite{Bal}). Another system where the finiteness is  at the
forefront of the theoretical investigation since several decades
is the superfluid atomic nucleus. As a matter of fact many of the
theoretical tools such as particle number projection, even-odd
effects, number parity, blocking effect, particle - particle
Random Phase Approximation (pp-RPA), etc. have first been
developed in nuclear physics \cite{RS} before finding their
application to finite systems of condensed matter. Also the
schematic pairing model with which we will mostly deal in this
paper, namely the Picket Fence Model (PFM), whose exact solution
has been found by Richardson and Sherman \cite{Rich}, has
essentially been developed in the context of nuclear physics for
the description of deformed superfluid nuclei. For finite
condensed matter systems an early theoretical description was
proposed by M\"uhlschlegel, Scalapino, and Denton \cite{SPA} using
the Static Path Approximation (SPA) to the partition function.
This work stayed rather singular for a long time but the SPA has
recently been applied successfully to the PFM both in the
condensed matter \cite{Fal} and nuclear \cite{Ros1,Ros2} contexts.
A further standard method to treat quantum fluctuations namely the
well known RPA has quite extensively been used for nuclear systems
\cite{RS,RPA1} but equally for condensed matter problems
\cite{MAX}.

In this work we will further elaborate on the RPA approach. We
indeed have recently had quite remarkable success with a self
consistent extension of the pp-RPA, which we called
Self-Consistent RPA (SCRPA), by reproducing very accurately
groundstate and excitation energies of the PFM \cite{DS1} at zero
temperature. This formalism was also developed independently by
R\"{o}pke and collaborators who called it Cluster - Hartree - Fock
(CHF) \cite{Rop}. Such type of generalization of the RPA theory
grew out of the works of K.-J. Hara and D. Rowe \cite{Row} several
decades ago. Shortly afterwards the theory was rederived using the
method of many body Green functions \cite{Sch}. The success of the
theory motivates us to develop the SCRPA formalism also for the
finite temperature case and to study the thermodynamic properties
of the BCS Hamiltonian using the PFM as an example. For the
extension of SCRPA to finite temperature we use the Matsubara
Green functions approach \cite{Fet}. It appeared that the
approximation scheme is very effective in treating two-body
correlations in the particle-particle (pp) channel as well as the
Pauli principle effects. We should mention that we will work with
real particles and not with quasiparticles what should limit our
approach to temperatures above the critical temperature $T_c$
(i.e. to the normal phase). However, as we will see below, the
definition of $T_c$ in SCRPA is not so clear and we will be able
to continue our calculation quite deeply into the superfluid
regime.

We organize the paper in the following way. In Section~2, the
approach is outlined in general. Then, in Section~3, the formalism
is applied to the PFM. A comparison with the exact solutions as
well as with the results of other approximations is made in
Section~4. Section 5 is devoted to comparison with other recent
works. In section 6, we will summarize the results and draw some
conclusions. In an appendix a variant for the calculation of the
occupation numbers is proposed.

\section{General formalism}

In treating a finite many-body system at finite temperature, it is
convenient to use the grand canonical ensemble although it violates the
number conservation. With the definition
\[
K=H-\mu N
\]
the grand partition function and statistical operator read
\[
Z_G=e^{-\beta \Omega }=Tr(e^{-\beta K})
\]
\[
\rho _{_G}=Z_G^{-1}e^{-\beta K}=e^{\beta (\Omega -K)}
\]
where $\beta =1/T.$ Then for any Schr\"{o}dinger operator $A_\alpha $ the
modified Heisenberg picture can be introduced
\[
A_\alpha (\tau )=e^{K\tau }A_\alpha e^{-K\tau }
\]
and the temperature (or the Matsubara) Green's Function (GF) is defined as \cite{Fet}
\[
G_{\alpha \beta }^{\tau -\tau ^{\prime }}=-\left\langle T_\tau A_\alpha
(\tau )A_\beta ^{+}(\tau ^{\prime })\right\rangle =-Tr\left[ e^{-\beta
(K-\Omega )}T_\tau e^{\tau K}A_\alpha e^{-(\tau -\tau ^{\prime })K}A_\beta
^{+}e^{-\tau ^{\prime }K}\right] =
\]
$$
=-Tr\left[ \rho _{_G}T_\tau e^{\tau K}A_\alpha e^{-(\tau -\tau ^{\prime
})K}A_\beta ^{+}e^{-\tau ^{\prime }K}\right] \eqno(1)
$$
Here, the brackets $\langle \phantom{O}\rangle $ mean the thermodynamic
average; $T_\tau $ is a $\tau $ ordering operator, which arranges operators
with the earliest $\tau $ (the closest to $-\beta $) to the right.

Let us consider the two-body Hamiltonian

$$
H=\sum_{12}t_{12}a_1^{+}a_2+\frac 14\sum_{1234}\overline{v}%
_{1234}a_1^{+}a_2^{+}a_3a_4 \eqno(2)
$$
where $a,\ a^{+}$ are fermion annihilation and creation operators; \medskip $%
t_{12}$ and $\overline{v}_{1234}=v_{1234}-v_{1243}$ are the
kinetic energy and the antisymmetrized matrix element of the
two-body interaction. The Green's function $G_{\alpha \beta
}^{\tau -\tau ^{\prime }}$ for an arbitrary operator $A_\alpha
^{+}$ obeys the following equation of motion:
\[
-\frac \partial {\partial \tau }G_{\alpha \beta }^{\tau -\tau ^{\prime
}}=\delta _{\tau -\tau ^{\prime }}\left\langle \left[ A_\alpha ,A_\beta ^{+}%
\right] \right\rangle -\left\langle T_\tau \left[ A_\alpha ,K\right] ^\tau
A_\beta ^{+}(\tau ^{\prime })\right\rangle =
\]
\[
=\delta _{\tau -\tau ^{\prime }}N_{\alpha \beta }+\sum_\gamma \int d\tau
_1^{\prime }{\cal H}_{\alpha \gamma }^{\tau -\tau _1^{\prime }}G_{\gamma
\beta }^{\tau _1^{\prime }-\tau ^{\prime }}
\]
In this expression it is possible to split the effective Hamiltonian ${\cal H%
}_{\alpha \beta }^{\tau -\tau _{}^{\prime }}$ into an
instantaneous and a dynamic (frequency dependent) part \cite{DRS}
\begin{eqnarray*}
{\cal H}_{\alpha \beta }^{\tau -\tau _{}^{\prime }} &=&\sum_{\beta ^{\prime
}}\left\{ \delta _{\tau -\tau ^{\prime }}\left\langle \left[ \left[ A_\alpha
,K\right] ,A_{\beta ^{\prime }}^{+}\right] \right\rangle -\left\langle
T_\tau \left[ A_\alpha ,K\right] ^\tau \left[ K,A_{\beta ^{\prime }}^{+}%
\right] ^{\tau ^{\prime }}\right\rangle _{irr}\right\} N_{\beta ^{\prime
}\beta }^{-1}= \\
&=&{\cal H}_{\alpha \beta }^{(0)}\delta _{\tau -\tau ^{\prime }}+ {\cal H}%
_{\alpha \beta }^{(r)\tau -\tau _{}^{\prime }}
\end{eqnarray*}
In the approximation of the instantaneous effective Hamiltonian
i.e. neglecting ${\cal H}_{\alpha \beta }^{(r)\tau -\tau
_{}^{\prime }}$, the Dyson equation for the two-body Matsubara GF
$G_{\alpha \beta }^{\tau -\tau ^{\prime }}$ can be written as
$$
-\frac \partial {\partial \tau }G_{\alpha \beta }^{(0)\tau -\tau ^{\prime
}}=\delta _{\tau -\tau ^{\prime }}N_{\alpha \beta }+\sum_\gamma {\cal H}%
_{\alpha \gamma }^{(0)}G_{\gamma \beta }^{(0)\tau -\tau ^{\prime }}\eqno(3)
$$

In the treatment of two particle correlations let us specify the arbitrary operator $%
A_\alpha $ as $A_{k_1k_2}=a_{k_1}a_{k_2}$. In this case the Dyson equation
(3) takes the following form in the frequency representation
$$
i\omega _nG_{k_1k_2k_1^{\prime }k_2^{\prime }}^{SCRPA}=N_{k_1k_2k_1^{\prime
}k_2^{\prime }}+\sum_{p_1p_2}{\cal H}_{k_1k_2p_1p_2}^{(0)}G_{p_1p_2k_1^{%
\prime }k_2^{\prime }}^{SCRPA}\,,\eqno(4)
$$
where, in supposing that the single particle density matrix is
diagonal in the basis used (this is for example the case
inhomogeneous matter):
$$
N_{k_1k_2k_1^{\prime }k_2^{\prime }}=\left\langle \left[
a_{k_1}a_{k_2},a_{k_{2^{\prime }}}^{+}a_{k_{1^{\prime }}}^{+}\right]
\right\rangle =\delta _{k_1k_2k_1^{\prime }k_2^{\prime }}\left(
1-n_{k_{2^{\prime }}}-n_{k_{1^{\prime }}}\right) \eqno(5)
$$
and
$$
{\cal H}_{k_1k_2k_1^{\prime }k_2^{\prime }}^{(0)}=\sum_{p_1p_2}\left\langle %
\left[ \left[ a_{k_1}a_{k_2},K\right] ,a_{p_2}^{+}a_{p_1}^{+}\right]
\right\rangle N_{p_1p_2k_1^{\prime }k_2^{\prime }}^{-1}\eqno(6)
$$
Here $\delta _{k_1k_2k_1^{\prime }k_2^{\prime }}$ is the antisymmetrized
Kronecker symbol and $n_k=\left\langle a_k^{+}a_k\right\rangle $ are the
single particle occupation numbers which can be found from the
single-particle Matsubara GF

\[
G_{kk^{\prime }}^\tau =\langle T_\tau a_k(\tau )a_{k^{\prime
}}^{+}(0)\rangle
\]
as
$$
n_k=\langle a_k^{+}a_k\rangle =\lim\limits_{\tau \rightarrow 0^{-}}G_k^\tau %
\eqno(7)
$$
In general the single-particle Matsubara Green's function $G_k^{\tau -\tau
^{\prime }}$ obeys the following Dyson equation:
$$
\left( -\frac \partial {\partial \tau }+\varepsilon _k\right) G_{kk^{\prime
}}^\tau =\delta (\tau )+\int d\tau _1M_k^{\tau -\tau _1}G_k^{\tau _1}\eqno%
(8)
$$
or in the frequency representation
$$
G_k^{\omega _n}=G_k^0+G_k^0M_kG_k^{\omega _n}\eqno(9)
$$
where
$$
G_k^0=\frac 1{i\omega _n-\varepsilon _k}\eqno(10)
$$
Here $\varepsilon _k$ contains already the usual (instantaneous)
mean field so that $M_k$ denotes only the dynamical part of the
mass operator.

Now the problem is to find an approximation for the mass operator
$M_k$ consistent with the SCRPA. A solution to this problem has
been proposed in \cite{DRS}, which goes via the two body $T$ --
matrix representation of the single particle mass operator
\cite{Mig}, evaluating $T$ within SCRPA.

Let us add at this point a word of physical interpretation of the
mean-field operator (6) \cite{DS1}. A quick look allows to realize
that it contains no higher than two body correlation functions and
therefore for their determination, with (4), one obtains a
selfconsistency problem. Furthermore one can consider the nucleus
as a gas of zero point pair fluctuations. These fluctuations
create their own mean field, i.e. one pair fluctuation moves in
the average potential created by all the other pair fluctuations.
This average pair fluctuation field is graphically represented in
Fig. 1. It gives rise, as usual, to a nonlinear problem. Of
course, the single particle mean field introduced in (9) and
further developed below is coupled to the selfconsistent pair
potential. This is the deeper meaning of ${\cal H}^{(0)}$ of (6).

\begin{figure}[tph]
\centering
\includegraphics[scale=0.6]{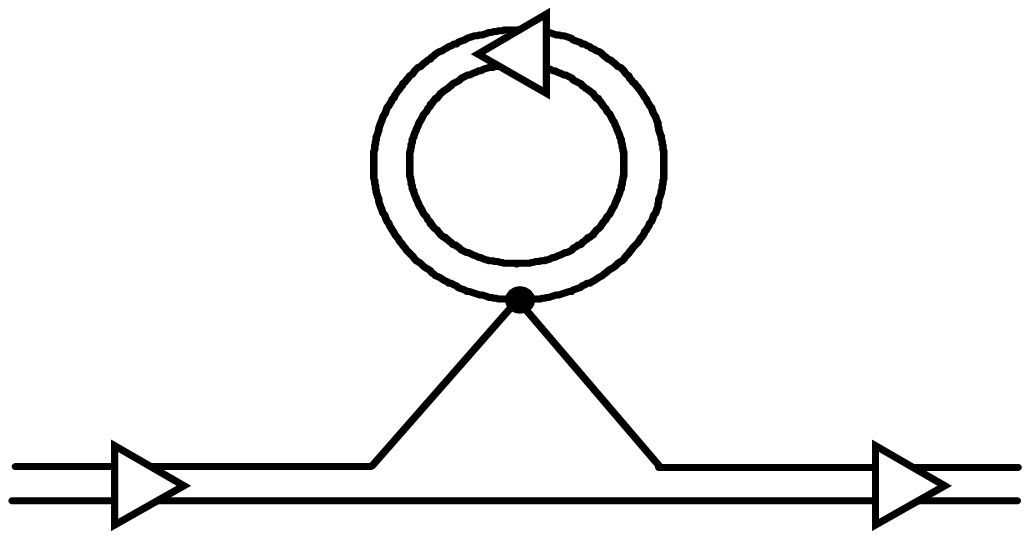}
\vspace{0.5cm} \caption{Schematic representation of the first
order self-energy of a pair fluctuation. Exchange terms are not
presented. The full black dot is the interaction.}
\end{figure}


\section{Application to the Picket Fence Model}

The model consists of an equidistant multilevel pairing
Hamiltonian with each level two fold degenerate, i.e. only spin
up/down fermions of one kind can occupy one level. The
corresponding Hamiltonian is given by
$$
H=\sum_{k=1}^\Omega e_kN_k-G\sum_{i,k=1}^\Omega P_i^{+}P_k\eqno(11)
$$
with
$$
N_k=c_k^{+}c_k+c_{\overline{k}}^{+}c_{\overline{k}},
$$
$$
P_k^{+}=c_k^{+}c_{\overline{k}}^{+}\eqno(12)
$$
where $\overline{k}$ means the time reversed of $k$, single
particle energies are $e_k=k\varepsilon -\lambda ,$ with level
spacing $\varepsilon $ chosen to be equal to $1$, and $\Omega $
stands for the number of levels. The chemical potential $\lambda $
will be chosen such as to conserve the average number of particles
$N=\Omega$ of the system. The operators defined in (12) form an
SU(2) algebra for each level $j$ and obey the following
commutation relations
$$
\left[ P_j, P_k^{+}\right]= \delta _{jk}(1-N_j),
$$
$$
\left[ P_j, N_k\right]= 2\delta _{jk}P_j,\eqno(13)
$$

\subsection{SCRPA equations}

To study the model at finite temperature we define in analogy to (1)
the following set of two-body Matsubara GFs
\[
G_{ji}^\tau =-\left\langle T_\tau \overline{P}_j(\tau )
\overline{P}_i^{+}(0)\right\rangle ,
\]
where
$$
\overline{P}_j=\frac{P_j}{\sqrt{<\left| 1-N_j \right|>}}
$$
Applying the instantaneous approximation for the mass operator we
obtain the expressions for the two body SCRPA GF's:

$$
i\omega _nG_{ji}^{SCRPA}=\delta _{ji}+\sum_k{\cal H}%
_{jk}^{(0)}G_{ki}^{SCRPA} ,\eqno(14)
$$
with
$$
{\cal H}_{jk}^{(0)}=2\delta _{jk}\left( e_j+\frac
G{<1-N_j>}\sum_{j^{\prime }}<P_j^{+}P_{j^{\prime }}>\right)
-G\frac{<(1-N_j)(1-N_k)>} {\sqrt{\left| <1-N_j><1-N_k>
\right|}}\eqno(15)
$$

To find the correlation functions of the form $<(1-N_j)(1-N_k)>$
we will use the following approximation:

\noindent when $j \neq k$ it is a simple factorization procedure,
which has turned out to be accurate in the zero temperature limit:
$$
<(1-N_j)(1-N_k)>= <1-N_j><1-N_k> \eqno(16a);
$$
but when $j=k$ we use the following exact relation
$$
<(1-N_j)^2>=<1-N_j>+2<P_j^{+}P_j>\eqno(16b)
$$
which can easily be obtained taking into account that $n_j^2=n_j$
and $n_jn_{\overline{j}}=P_j^{+}P_j$ (here $n_j=c_j^{+}c_j$). It
should be noted that in \cite{DS1} the factorization (16a) was
also used for the diagonal part (16b) and quite accurate results
were obtained. We will show below that with (16b) one obtains
still improved results. As shown in \cite{DS1} it is possible to
avoid above approximation. However, this is at the cost of a
considerable numerical complication. We refrain from this here
because it brings only very little improvement of results.

With this ansatz a particle-particle RPA-like equation is obtained
$$
G_{ji}^{SCRPA} =\delta
_{ji}\frac{1}{z-C_j}-\frac{G\sqrt{\left|D_jD_i\right|}}{\left(
z-C_j\right) \left( z-C_i\right) }\times \left[ 1+G\sum_k\frac{D_k}{z-C_k}%
\right] ^{-1} , \eqno(17)
$$
where
$$
z=i\omega _n,
$$
$$
D_i=<1-N_i>
$$
and
\[
C_j=2\left( e_j-Gn_j+\frac G{D_j}\sum_{j^{\prime }\neq j}<P_j^{+}P_{j^{\prime }}>
\right)
\]
From this one easily can find the excitation spectrum of the model in equating
the denominator of (17) to zero
$$
1+G\sum_k\frac{D_k}{z-C_k}=0 \eqno(18)
$$
Knowing the poles of the Green's function (17), one can write down
its spectral representation (we here give it as a function of
imaginary time), with the corresponding residua:
\[
-G_{p_1p_2}^\tau =\theta \left( \tau \right) \sqrt{D_{p_1}D_{p_2}}\left[
X_{p_1}^\mu X_{p_2}^\mu e^{-E_\mu \tau }\left( 1+n_B(E_\mu )\right)
+Y_{p_1}^\mu Y_{p_2}^\mu e^{E_\mu \tau }n_B(E_\mu )\right] +
\]

$$
+\theta \left( -\tau \right) \sqrt{D_{p_1}D_{p_2}}\left[ X_{p_1}^\mu X_{p_2}^\mu
e^{-E_\mu \tau }n_B(E_\mu )+Y_{p_1}^\mu Y_{p_2}^\mu e^{E_\mu \tau }\left(
1+n_B(E_\mu )\right) \right] \eqno(19a)
$$

\[
-G_{h_1h_2}^\tau =\theta \left( \tau \right) \sqrt{D_{h_1}D_{h_2}}\left[
Y_{h_1}^\mu Y_{h_2}^\mu e^{-E_\mu \tau }\left( 1+n_B(E_\mu )\right)
+X_{h_1}^\mu X_{h_2}^\mu e^{E_\mu \tau }n_B(E_\mu )\right] +
\]

$$
+\theta \left( -\tau \right) \sqrt{D_{h_1}D_{h_2}}\left[ Y_{h_1}^\mu Y_{h_2}^\mu
e^{-E_\mu \tau }n_B(E_\mu )+X_{h_1}^\mu X_{h_2}^\mu e^{E_\mu \tau }\left(
1+n_B(E_\mu )\right) \right] \eqno(19b)
$$

\[
-G_{ph}^{\tau }=-G_{hp}^{\tau }=\theta \left( \tau \right)
\sqrt{\left|D_{p}D_{h}\right|}\left[
X_{p}^{\mu }Y_{h}^{\mu }e^{-E_{\mu }\tau }\left( 1+n_{B}(E_{\mu })\right)
+Y_{p}^{\mu }X_{h}^{\mu }e^{E_{\mu }\tau }n_{B}(E_{\mu })\right] +
\]

$$
+\theta \left( -\tau \right) \sqrt{\left|D_{p}D_{h}\right|}
\left[ X_p^\mu Y_h^\mu e^{-E_\mu \tau
}n_B(E_\mu )+Y_p^\mu X_h^\mu e^{E_\mu \tau }\left( 1+n_B(E_\mu )\right) %
\right] \eqno(19c)
$$

\[
n_{B}\left( E_\mu \right) =\frac{1}{e^{\beta E_\mu}-1},
\]
where the index $p$ refers to the states above Fermi level and the
index $h$ to the ones below. The following amplitudes were
introduced in these formulas:
$$
X_{p}^{\mu }=\frac{\sqrt{GD_p}}{\left| C_{p}\right| -E_{\mu }}F_{\mu
}, \, \, Y_{h}^{\mu }=\frac{\sqrt{-GD_h}}{\left| C_{h}\right| +E_{\mu }}F_{\mu }\eqno%
(20a)
$$
$$
X_{h}^{\mu }=\frac{\sqrt{-GD_h}}{\left| C_{h}\right| -E_{\mu }}F_{\mu
}, \, \,Y_{p}^{\mu }=\frac{\sqrt{GD_p}}{\left| C_{p}\right| +E_{\mu }}F_{\mu }\eqno%
(20b)
$$
with
$$
F_\mu ^{-2}=\frac{\partial}{\partial z}\left[ 1+G\sum_k\frac{D_k}{z-C_k}\right] _{z=E_\mu }%
\eqno(20c)
$$
These amplitudes obey the usual normalization conditions
$$
\sum_{p}{}X_{p}^{\mu }X_{p}^{\mu ^{\prime }}+\sum_{h}{} Y_{h}^{\mu
}Y_{h}^{\mu ^{\prime }}=\delta _{\mu \mu ^{\prime }}
$$
$$
\sum_{h}{}X_{h}^{\mu }X_{h}^{\mu ^{\prime }}+\sum_{p}{}
Y_{p}^{\mu}Y_{p}^{\mu ^{\prime }}=-\delta _{\mu \mu ^{\prime
}}\eqno(21)
$$

Two-body correlation functions can be obtained from the Green's
function (19) as follows
\[
<P_{p_1}^{+}P_{p_2}>=-G_{p_1p_2}^{\tau \rightarrow
0^{-}}=\sqrt{D_{p_1}D_{p_2}}\left( \sum_\mu X_{p_1}^\mu X_{p_2}^\mu n_B(E_\mu
)+\sum_\mu Y_{p_1}^\mu Y_{p_2}^\mu \left( 1+n_B(E_\mu )\right) \right)
\]

$$
<P_{h_1}^{+}P_{h_2}>=-G_{h_1h_2}^{\tau \rightarrow
0^{-}}=\sqrt{D_{h_1}D_{h_2}}\left( \sum_\mu Y_{h_1}^\mu Y_{h_2}^\mu n_B(E_\mu
)+\sum_\mu X_{h_1}^\mu X_{h_2}^\mu \left( 1+n_B(E_\mu )\right) \right) \eqno%
(22)
$$

\[
<P_p^{+}P_h>=-G_{ph}^{\tau \rightarrow
0^{-}}=\sqrt{\left| D_pD_h\right|} \left( \sum_\mu X_p^\mu Y_h^\mu n_B(E_\mu
)+\sum_\mu Y_p^\mu X_h^\mu \left( 1+n_B(E_\mu )\right) \right)
\]

\subsection{Occupation numbers in the SCRPA}

In order to close the set of the SCRPA equations, it is necessary
to find the so far unknown occupation numbers  $n_k=<c_k^+c_k>$ .
For this, we should find the single particle Green's function
$G_{k}^\tau$ consistent with the SCRPA scheme. As discussed in
sect. II, the single particle mass operator $M_k$ has in general
the exact representation in terms of the two body $T$-matrix
\cite{Mig} and then an appropriate approximation for the
$G_{k}^\tau$ can be obtained. It consists in using the mass
operator $\widetilde {M_k}$ calculated through the $T$-matrix
found in the framework of SCRPA. As the relation between the
$T$-matrix and the sum of the all irreducible Feynman graphs in
the pp-chanel is also known then the following expression for the
single particle mass operator can be obtained \cite{DRS}:
$$
\widetilde{M_k}= G\sum_{k_1k_2}G_{\overline{k}}^{0(\tau_1^{\prime }-\tau_1)}
G_{k_1k_2}^{(\tau_1^{\prime }-\tau_1)}\widetilde{{\cal H}}%
_{k_2k}^{(0)}\eqno(23)
$$
$\widetilde{\cal H}_{k_2k}^{(0)}$ is expressed through the
effective Hamiltonian (15) without the disconnected part
$$
\widetilde{\cal H}_{kk\prime}^{(0)}={\cal
H}_{kk\prime}^{(0)}-2\delta_{kk\prime} \varepsilon_{k} \eqno(24)
$$
where $\varepsilon_k$ is defined below in (30).

In addition to this transparent scheme there also exists an additional
consistency requirement \cite{DS2}.
It follows from the possibility to calculate the average value of the
Hamiltonian $<H>$ in two ways.
On the one hand one has the following relation between the single particle Green's
function (9) and  $<H>$ \cite{Fet}:
$$
<H> =-\frac{1}{2}\lim_{\tau^{\prime }-\tau\rightarrow 0^{+}}
\sum_k\left[ \frac \partial {\partial \tau}-e_k\right] \left(
G_k^{(\tau-\tau^{\prime })}+G_{\overline{k}}^{(\tau-\tau^{\prime })}\right)
\eqno(25)
$$
On the other hand there exists the straightforward calculation of
 $<H>$ through the two body Green's functions (22)
$$
<H>=\sum_ke_k<N_k>-G\sum_{k_1 k_2}<P^+_{k_1}P_{k_2}>
$$
$$
=-\frac{N^2}{4}+2\sum_p e_p<N_p>-2\sum_p \sum_\mu \sqrt{GD_p}
F_\mu \left( n_B(E_\mu)X_p^\mu+(1+n_B(E_\mu))Y_p^\mu \right) \eqno (26)
$$
where we used the particle-hole symmetry of the model [11] and reduced sums over p and h
only to the one over the particle states.

The additional consistency condition lies in the requirement that
both expressions (25) and (26) should give exactly the same
results. This only is satisfied if the single particle GF is
expanded to first order in the renormalized single particle mass
operator (23) (one may verify that this is in analogy to the
standard RPA scheme, i.e. the standard RPA average energy is
obtained via (25) using a single particle GF with only
perturbative renormalization from the RPA-modes):
$$
G_p=G_p^0+G_p^0 M^{SCRPA}_p G_p^0 \eqno(27)
$$
with
$$
M^{SCRPA}_p=\frac{\sqrt{D_p}}{D_p^0}\widetilde{M_p}, \eqno(28)
$$
$$
D_p^0=1-f_p-f_{\overline p}
$$
$$
f_p=\frac{1}{1+e^{\varepsilon_p \beta}}, \eqno(29)
$$
and
$$
\varepsilon_p=e_p-Gf_p\frac{f_p}{n_p}\frac{D_p}{D_p^0}. \eqno(30)
$$

Finally we get the following expression for the SCRPA single particle mass operator
$M^{SCRPA}_p$:
$$
M_{p}^{SCRPA} =\frac{\sqrt{GD_p}}{D_p^0}\sum_{\mu }F_{\mu}\left[
\frac{S^{(1)}_{\mu p}(f_{p}+n_B(E_\mu )) } {i\omega _n+\varepsilon _p-E_\mu }+
\frac{S^{(2)}_{\mu p}(1-f_{p}+n_B(E_\mu )) } {i\omega _n+\varepsilon _p+E_\mu }
\right] \
\eqno(31)
$$
where
$$
S^{(1)}_{\mu p}=-\sum_{p^{\prime }}X_{p^{\prime }}^\mu
\widetilde{\cal H}_{pp^{\prime }}+\sum_{h^{\prime }}Y_{h^{\prime }}^\mu
\widetilde{\cal H}_{ph^{\prime }},
$$
$$
S^{(2)}_{\mu p}=-\sum_{p^{\prime }}Y_{p^{\prime }}^\mu
\widetilde{\cal H}_{pp^{\prime }}+\sum_{h^{\prime }}X_{h^{\prime }}^\mu
\widetilde{\cal H}_{ph^{\prime }}
$$

The corresponding single particle occupation numbers $n_p$, found from (7),
is the following
$$
n_{p}=<c_{p}^{+}c_{p}>=f_p+\frac{\sqrt{GD_p}}{D_p^0}\sum_{\mu }F_{\mu}
\left[ X_p^{\mu}\frac{ n_{B}(E_{\mu })D_p^0-f_p^2}{\left( 2\varepsilon _{p}-E_{\mu }\right)}
+
Y_p^{\mu}\frac{ (n_{B}(E_{\mu })+1) D_p^0+f_p^2}{\left( 2\varepsilon _{p}+E_{\mu }\right)}
\right]
$$
$$
-f_p \left( 1-f_p \right) \beta \frac{\sqrt{GD_p}}{D_p^0}\sum_{\mu} F_{\mu}
\left[X_p^{\mu}(f_p+n_{B}(E_{\mu })+Y_p^{\mu}(1-f_p+n_{B}(E_{\mu }) \right] \eqno(32)
$$

Let us demonstrate now that using the single particle Green's
function (27) with the mass operator (31) and occupation numbers
(32) in the calculation of the average energy (25) indeed leads to
the equation (26). At first one finds the derivative of the single
particle GF (27)
$$
\frac {\partial G_p^{(\tau-\tau^{\prime })\rightarrow 0^{+}}} {\partial \tau}=
-\varepsilon_p f_p+\frac{\sqrt{GD_p}}{D_p^0}\sum_\mu F_\mu \left \{ (1-f_p)
\left[ S_{\mu p}^{(1)}\frac{(\varepsilon_{p}-E_\mu)n_B(E_\mu)}{(2\varepsilon _{p}-E_\mu)^2}
+S_{\mu p}^{(2)} \frac{(\varepsilon_{p}+E_\mu)(1+n_B(E_\mu))}{(2\varepsilon _{p}+E_\mu)^2}
\right] \right.
$$
$$
+\varepsilon_{p}f_p
\left[S_{\mu p}^{(1)}\frac{f_p+n_B(E_\mu)}{(2\varepsilon _{p}-E_\mu)^2}
+S_{\mu p}^{(2)} \frac{1-f_p+n_B(E_\mu)}{(2\varepsilon _{p}+E_\mu)^2}
\right]
$$
$$
+\beta\varepsilon_{p}f_p(1-f_p)
\left[ S_{\mu p}^{(1)}\frac{f_p+n_B(E_\mu)}{2\varepsilon _{p}-E_\mu}
+S_{\mu p}^{(2)} \frac{1-f_p+n_B(E_\mu)}{2\varepsilon _{p}+E_\mu}
\right]
$$
$$
-\left. f_p\left[ S_{\mu p}^{(1)}\frac{f_p+n_B(E_\mu)}{2\varepsilon _{p}-E_\mu}
+S_{\mu p}^{(2)} \frac{1-f_p+n_B(E_\mu)}{2\varepsilon _{p}+E_\mu}
\right]  \right\}=
$$
$$
=-\varepsilon_{p}n_p+\sum_\mu \frac{\sqrt{GD_p}F_\mu}{D_p^0}
\left[ S_{\mu p}^{(1)}\frac{n_B(E_\mu)D_p^0-f_p^2}{2\varepsilon _{p}-E_\mu}
+S_{\mu p}^{(2)} \frac{n_B(E_\mu)D_p^0+f_p^2}{2\varepsilon _{p}+E_\mu}
\right]
$$
Inserting this expression in (25) we obtain
$$
<H> =-\frac{1}{2}\lim_{\tau^{\prime }-\tau\rightarrow 0^{+}}
\sum_k\left[ \frac \partial {\partial \tau}-e_k\right] \left(
G_k^{(\tau-\tau^{\prime })}+G_{\overline{k}}^{(\tau-\tau^{\prime })}\right)=
$$
$$
=-\frac{N^2}{4}+2\sum_p (e_p+\varepsilon_{p})n_p-2\sum_{p \mu}
\sqrt{GD_p}F_\mu\left(n_B(E_\mu)X_p^\mu + (1+n_B(E_\mu))Y_p^\mu\right)
+G\sum_p \frac{D_p}{D_p^0}f_p^2=
$$
$$
=-\frac{N^2}{4}+4\sum_p e_p n_p-2\sum_p \sum_\mu \sqrt{GD_p}
F_\mu \left( n_B(E_\mu)X_p^\mu+(1+n_B(E_\mu))Y_p^\mu \right)
$$
This is exactly equal to (26).

The system of the SCRPA equations is fully closed now. Together
with (17), (19) and (22) this represents a self-consistent problem
for pair fluctuations.

We want to indicate at this point that the above way to determine
the single particle occupancies is not the only possibility. In
the Appendix we will give another variant which, however, yields
results close to the ones with the method of this section. The non
uniqueness of the occupation numbers reflects the fact that with
the truncated  ansatz (3), at zero temperature, no corresponding
ground state wave function can be found, as explained in
\cite{DS1}. For a wave function to exist, the ansatz (3) must be
extended. It can, however, be shown that the correction terms are
small \cite{new}.

\subsection{Exact statistical treatment of the PFM}

For an exact statistical treatment of the Picket Fence Model we
have to find all exact eigenvalues and eigenstates of the
Hamiltonian (11). Since singly-occupied levels do not participate
in the pair scattering, eigenstates can be classified according to
the number of unpaired particles $S$ (seniority). There are
$C_{\Omega-S}^{\Omega}=\frac{\Omega!}{S!(\Omega-S)!}$ different
multiplets of this type, each of dimension
$C_{\frac{N-S}{2}}^{\Omega-S}$ and degeneracy $2^S$. If we define
the following set of basis states for each multiplet:
$$
|\left\{ s_{i},N_{i}\right\}>\eqno(33)
$$
where $s_{i}=N_{i}=1$ for singly-occupied levels and $s_{i}=0,
N_{i}=0$ or $2$ for remaining levels ($\sum_i s_i=S$), the
Hamiltonian matrix will have the following diagonal and
off-diagonal elements
$$
\left\langle \left\{ s_{i},N_{i}\right\}\left| H\right| \left\{
s_{i},N_{i}\right\} \right\rangle =\sum_{k\in S}\left(
e_{k}-\lambda \right) +\sum_{k\in \Omega-S}\left[ \left(
e_{k}-\lambda \right) -\frac{G}{4}\left( 4-N_{k}\right) \right]
N_{k}\eqno(34)
$$
$$
\left\langle s_{j_{1}}N_{j_{1}},...,s_{j_{k}}2,...,s_{j_{i}}0,...,s_{j_{n}}N_{j_{n}}%
\left| H\right| s_{j_{1}}N_{j_{1}},...,s_{j_{k}}0,...,s_{j_{i}}2,...,s_{j_{n}}N_{j_{n}}%
\right\rangle =-G\eqno(35)
$$
The exact eigenvalues and eigenstates can be calculated by
diagonalization of this matrix in each multiplet. The exact grand
canonical average $<A>$ of any operator can then be obtained with
the help of the grand partition function $Z_G=Tre^{-\beta
\hat{H}}$ and the statistical operator
$\hat{\rho_G}=Z_G^{-1}e^{-\beta \hat{H}}$ as
$$
<A>=Tr[\hat{A}\hat{\rho_G}]\eqno(36)
$$

\section{Results and Discussion}

In order to check the accuracy of our theory and of the different
approximations schemes we first calculate the average energy of
the system $<H>$ as a function of the particle number $N$ and
temperature $T$. The results of the different calculations are
presented in Figures 2 -- 4. Calculations were made for
a value of the pairing constant $G$ which is smaller but close to the critical value $%
G_{cr}$ at $T=0$. The phase transition from the normal to the
superfluid phase occurs in the system when $G \to G_{cr}$ . We
compare the SCRPA results with the exact ones for the Grand
Canonical Ensemble (GCE) as well as results of the standard
Thermal RPA (TRPA) and Thermal MFA (TMFA). One can see that when
the number of levels $\Omega $ (and number of particles $N$)
increases the description of the intrinsic energy becomes better and at $%
\Omega = N = 10$ the TSCRPA results practically coincide with the
exact ones. Especially the last case will be considered below more
carefully.

Let us now come to the discussion of the behavior of the
excitation energies. The dependence of the excitation energies of
the addition mode (see \cite{DS1}) as a function of $G$ is shown
in Fig. 5 at zero temperature.  The SCRPA (solid lines) is
compared with the standard RPA calculations (dashed lines) and the
exact ones (open circles). Increasing the interaction constant,
the lowest energy in the RPA goes to zero and at $G_{cr} \simeq
0.33$ the collapse takes place which is connected with the
transition from the normal to the superfluid phase. At finite
temperature (see Fig. 6 where the dependence of the lowest
addition mode is presented as a function of $G$ at $T=0.5$) this
collapse occurs at a higher value of the interaction constant
($G_{cr} \simeq 0.43$) what is due to the reduced intensity of the
residual interaction because of the thermal factors. This collapse
is absent in the exact calculations at zero temperature and also
in the SCRPA calculations at zero and finite temperatures. It is
also remarkable that the SCRPA yields a rise of all excitation
energies with increasing $G$ in contrast with RPA and in very good
agreement with the exact results. This comes from the fact that in
the PFM with the Kramer's degeneracy of levels the Pauli repulsion
is extremely strong overruling the original attractive
interaction. In this model, therefore, standard RPA gives
qualitatively wrong result.

We next consider the behavior of the system near the phase transition point.
To make  distinctions between different
results more apparent we not only show the full intrinsic energy $<H>$ but also
the correlation energy $E_{corr}$ which is defined as
$$
E_{corr}=<H>-<H>_0\eqno(37)
$$
where $<H>_0$ is the average energy calculated in Mean Field
Approximation.
In Figures 7 and 8, the average energy $<H>$ and correlation
energy $E_{corr}$ as a function of $T$ are displayed for the
interaction constant $G=0.4$ (at $T=0$ this value of $G$ is larger than $%
G_{cr} \simeq 0.33$). With increasing $T$ the mean field
rearrangement occurs and the system goes from the superfluid phase
to the normal one at $T_c\simeq 0.38$. Note, that within the TRPA
the lowest excitation energy $\omega _1$ vanishes when
$T\rightarrow T_c$, whereas within the TSCRPA $\omega _1$ stays
finite. For both the correlation energy and the intrinsic energy
the TSCRPA gives more precise results as compared to the other
approximations. It is remarkable that the TSCRPA results are
accurate down to practical zero temperature, in spite of the fact
that within standard BCS theory one enters the superfluid regime.
A quasiparticle formulation of SCRPA \cite{SCQRPA} will only be
necessary for stronger $G$ values driving the system more deeply
into the symmetry broken phase.

To analyze the region near the phase transition point in more detail, the
heat capacity is calculated as a partial derivative of
the intrinsic energy with respect to $T$
$$
C_v=\frac{\partial <H>}{\partial T}\eqno(38)
$$
The results
are shown in Fig.~9. The TRPA and TMFA give discontinuities of $C_v$ at $%
T_c$ (we recall again that our results are obtained using a normal
fluid approach and not transforming to quasiparticles). The heat
capacity calculated in the TSCRPA has some kink near $T_c$ but has
no discontinuities and is quite similar to the exact result
through out the whole range of temperature.

Nevertheless, the TSCRPA and also the exact solution feel the
phase transition to the superfluid phase. It can be seen in Fig.~8
where both the TSCRPA and the exact correlation energies show a
depression near $T=0.38$. This originates from strong pair
fluctuations leading to the BCS phase transition in TMFA with the
critical temperature $T_{cr}^{BCS}=0.38$ for $G=0.4$. However, one
notices (see also \cite{duc-s}) that the sharp phase transition of
mean field is in reality completely smeared out and only a faint,
though clearly visible, signal survives.

To investigate the formation of such fluctuations in more detail,
it is useful to consider the spectral function \cite{Fet}
$$
A(k,\omega )=\frac{1}{i}\left[
G_k^{\omega_n}\biggl|_{i\omega_n\rightarrow \omega-i\eta}-
G_k^{\omega_n}\biggr|_{i\omega_n\rightarrow \omega+i\eta}
\right]\eqno(39)
$$
which includes two-body correlations through the self-consistent
treatment of the mass operator $M_k^{SCRPA}(\omega_n)$. Based on
the spectral function the density of states can be calculated as
\cite{SRS}
$$
 N(\omega )=\sum_k A(k,\omega )\eqno(40),
$$
The results of the calculations of $N(\omega )$ with $G_k$ of (27)
in (39) are shown in Fig.~10 at different values of $T$ and for
$G$ = 0.4. It is clearly seen that the distance between the two
quasiparticle peaks around the Fermi energy $\varepsilon_F$
($\omega=0$) increases with decreasing temperature. This process
sets in even above the BCS transition temperature
$T_c^{BCS}=0.38$. This rarefaction of the level density around
$\varepsilon_F$ above $T_c$ is not avoid of similarity with the
situation in high $T_c$ -- superconductors where a so-called
'pseudo gap' in the level density appears already above $T_c$
\cite{Tcr}. This 'pseudo gap' also is often attributed to a
decrease in the level density around $\varepsilon_F$ due to pair
fluctuations \cite{SRS,Tcr2}. Apparently it is a quite generic
feature that pair correlations diminish the density of levels
around $\varepsilon_F$ whereas particle-hole correlations give
rise to an increase.

In order to make the temperature dependence of the gap more
transparent let us introduce an effective (or canonical) gap which
recently was proposed in Eq. (22) of Ref. \cite{Del2}
$$
\Delta=G\sqrt{\sum_{ik}\left( \langle P_i^{+}P_k \rangle -
\left\langle
c_{i}^{+}c_{k}\right\rangle \left\langle c_{\overline{i}}^{+}c_{\overline{k}%
}\right\rangle \right) }\eqno(41)
$$
In the BCS approximation, the effective gap $\Delta$ coincides
with the usual grand canonical BCS gap
$$
\Delta_0=G\sum_{k}\langle P_k \rangle_{BCS}
$$
The dependencies of the effective gap on the interaction constant
$G$ at zero temperature and on temperature $T$ at $G=0.4$ are
shown in Figures 11 and 12. The SCRPA results give a very good
description of the gap at zero and non zero temperatures. It is
clearly seen that the SCRPA and exact calculations do not display
the phase transition at the point where the BSC gap vanishes.
Notice that in Fig. 11 the SCRPA result deteriorates for values of
$G$ well beyond the critical value. In this regime a quasiparticle
generalization of the SCRPA is necessary \cite{SCQRPA} because one
enters deeply in the superfluid region.

\section{Comparison with other works}

The PFM has recently widely been used for the study of quantum
pair fluctuations at finite temperature both in the context of
nuclear physics \cite{Ros1,Ros2} and of ultrasmall metallic grains
\cite{Fal,Bal}. In both fields the SPA approach where one
additionally takes into account number parity projection and
quantum (RPA) fluctuation around mean field was employed.

To compare our results with the above mentioned formalism let us
introduce the relevant energy scales. These are the average level
spacing $\delta$ and the BCS energy $\Delta=\Omega \delta
/[2\sinh(\delta / G)]$ \cite{Del1,Del2}. The properties of the
system described by the pairing Hamiltonian can be calculated as
universal functions of the single scaling parameter $\delta /
\Delta$. As long as the grain is not too small $\delta << \Delta$,
the fluctuation region around $T_c$ is narrow, and the mean field
(BCS) description of superconductivity is appropriate. It is the
case in ref. \cite{Ros2}, where the PFM is investigated with
characteristic values of $\delta / \Delta \simeq 0.2$. The result
of that work shows that the domain where the parity projection is
important lies in a small region near $T_c$. At temperatures
higher than $T_c$ the role of the fluctuation is decreased and
usual SPA becomes rather good in reproducing the exact canonical
results.

When the size of the system is decreased, fluctuations start to
smear the normal -- superconducting transition. The finite level
spacing suppresses the BCS gap and when $\delta$ becomes of the
order of $\Delta$, the fluctuation region becomes of order $T_c$
and the BCS description of superconductivity breaks down even at
zero temperature. Returning to our calculations, we can see that
SCRPA yields the best results for $\delta / \Delta \geq 1$ (see
Fig. 11, 12). This region corresponds to ultrasmall grains where
strong pairing fluctuations are dominant. In this sense our
results can be compared (at least qualitatively) with the results
of \cite{Fal}. To make this comparison more accurate we calculated
average energy and specific heat for a system with 50 fermions and
interaction constant $G=0.127$ and $G=0.255$, what corresponds to
$\delta / \Delta = 50$ and $\delta / \Delta = 1$ respectively. In
general our results correlate well with \cite{Fal}. From Fig. 13,
where $C_v$ is displayed as a function of $T/\delta$, we can see
that when $\delta / \Delta \to \infty$ the specific heat
approaches a linear dependence with temperature while when $\delta
\sim \Delta$ a bump structure arises at low temperature which is a
sign of the presence of strong pairing fluctuations. As it has
been seen from our previous discussion, the SCRPA gives a better
description with increasing particles number. And while we did not
perform exact GCE calculations for the system with 50 particles,
we can expect from our studies above and in \cite{DS1} that our
result should be very close to the exact one.

In conclusion of this section we can say that the results of
TSCRPA are at least as good as the ones of SPA with extensions.
However, contrary to SPA, TSCRPA has no problem at low temperature
and excitation energies and correlation functions can be
calculated directly as a function of temperature.

\section{Conclusion}

In this work we generalized our recent work \cite{DS1} on the
multilevel pairing model (PFM) within the SCRPA approach to finite
temperature (TSCRPA). The PFM has been recognised to account in
many respects for the physics of (superconducting) metallic
nano-grains. In our context SCRPA can be viewed as a
self-consistent mean field theory for pair fluctuatuations. The
results at $T=0$ in \cite{DS1} are in very close agreement with
the exact ones obtained from the Richardson procedure \cite{RS}.
It is therefore an important issue to also exploit SCRPA at finite
$T$ and to assess its accuracy with respect to exact results in
this case. Our comparison is mostly done for the case of ten
levels with $10$ particles where it is still of some ease to
establish the exact partition function. We, however, also
considered with TSCRPA the case of 50 particles in $50$ levels
assuming that the results are of equal quality or even better as
the ones obtained for $10$ particles. We base our studies on the
Matsubara $1$ and $2$ particle Green's functions which allows us
to calculate correlation and excitation energies, specific heat,
level densities, etc. It can be considered as a general advantage
of our approach that all these quantities are directly accessible
in the whole range of temperatures and coupling constants. For the
latter this holds in this work only true for interaction values
not driving the system deeply into the superfluid regime, since in
this work we only have been working with normal particles and not
with quasiparticles. For $G>>G_{cr}$ we have to employ the Self
Consistent Quasiparticle RPA (SCQRPA). It has recently been
demonstrated in the two level pairing model that also SCQRPA gives
very promising results \cite{SCQRPA}. The quality of our results
for the above mentioned quantities turns out to be excellent and
it does not fail in any qualitative nor quantitative aspect. Most
of the time the agreement with the exact solution is within the
couple of percent level. One particularly interesting feature of
our investigation is the fact that we achieved to calculate the
single particle Green's function consistently within TSCRPA. This
enabled us to give, for the first time, for the PFM the evolution
of the single particle level density with temperature. The
construction of the exact solution for this quantity is very
cumbersome and we refrained from working this out here. However,
backed with the positive experience for all other above mentioned
quantities we believe that also the level density is reasonably
accurate. The interesting aspect of our calculation is that with
decreasing temperature the density of single particle states
around the Fermi level decreases even above the critical
temperature as defined by BCS -- theory. It is suggestive to see
this feature in analogy to the appearance of a so called pseudogap
in high $T_c$ superconductors where also a depression in the level
density is observed approaching $T_c$ from above \cite{SRS,Tcr}.
It would be interesting to attempt an experimental verification
with metallic nanograins of our prediction that indeed the density
of levels rarefies with decreasing temperature already in the non
superfluid regime.

We also gave a short comparison of TSCRPA with results at finite
$T$ obtained with other approaches like the static path
approximation (SPA) to the partition function. Though the results
seem generally comparable, we think that TSCRPA is more versatile,
giving direct access to correlation functions, level densities,
excitation energies, etc. in the whole temperature range,
quantities which are otherwise difficult to obtain.

In the present work we restricted ourselves to values of the
coupling which are below or slightly above the critical value. In
the future we shall elaborate on the SCRPA for quasiparticles at
finite $T$ (TSCQRPA) which will allow us to consider the system
deeply in the superfluid phase and to study the transition from
one phase to the other in more detail.

\vspace{1cm}

\begin{center}

\textbf{ACKNOWLEDGMENTS}

\end{center}

\vspace{0.5cm}

We acknowledge useful discussions with P. Bozek, F. Hekking and J.
Hirsch.

\vspace{1cm}

\begin{center}

{\textbf{APPENDIX: OCCUPATION NUMBER -- VARIANT}}

\end{center}

 \vspace{0.5cm}

Let us give another possible way to find the occupation numbers at
finite temperature. Before to complete this task we firstly derive
the SCRPA equations with the Equation of Motion method and find
some useful relations between phonon amplitudes at zero
temperature. To attain this let us introduce the pair addition and
removal operators (phonons) \cite{DS1} as
$$
Q_\mu ^{+}=\sum_pX_p^\mu \overline{P}_p^{+}-\sum_hY_h^\mu
\overline{P}_h^{+}\eqno(A1)
$$
$$
Q_{\lambda }^{+}=\sum_{h}X_{h}^{\lambda }\overline{P}_{h}-
\sum_{p}Y_{p}^{\lambda }\overline{P}_{p} \eqno(A2)
$$
and apply the variational procedure where all expectation values
are found with respect to the vacuum of phonons (A1,A2).
$$
<\left[ \delta Q_\nu ,\left[ {\cal H},Q_\nu ^{+}\right] \right]
>=\omega _\nu <\left[ \delta Q_\nu ,Q_\nu ^{+}\right] > \eqno(A3)
$$
If we use the factorisation procedure (16) the following system of
equations for the phonon amplitudes is obtained
$$
\sum_{p^{\prime }}X_{p^{\prime }}^\mu A_{pp^{\prime
}}-\sum_{h^{\prime }}Y_{h^{\prime }}^\mu B_{ph^{\prime }}=E_\mu
X_p^\mu
$$
$$
\sum_{p^{\prime }}X_{p^{\prime }}^\mu B_{hp^{\prime
}}-\sum_{h^{\prime }}Y_{h^{\prime }}^\mu C_{hh^{\prime }}=-E_\mu
Y_h^\mu , \eqno(A4)
$$
where
$$
A_{pp^{\prime }}=2\delta _{pp^{\prime }} [ e_p] + 2G\delta
_{pp^{\prime }} \left[ \sum_{l\neq
p}\frac{<P_l^{+}P_p>}{D_p}-n_p\right] -G\sqrt{D_pD_{p^{\prime }}}
=2\delta _{pp^{\prime }}\varepsilon _p+\widetilde{A}_{pp^{\prime
}}
$$
$$
B_{ph}=-G\sqrt{-D_pD_h}
$$
$$
C_{hh^{\prime }}=2\delta _{hh^{\prime }}[ e_h-G] +
2G\delta _{hh^{\prime }}%
\left[ \sum_{l\neq h}\frac{<P_l^{+}P_h>}{D_h}+1-n_h\right]
-G\sqrt{D_hD_{h^{\prime }}} =-\left( 2\delta _{hh^{\prime
}}\varepsilon _h+\widetilde{C}_{hh^{\prime }}\right)
$$
$$
B_{hp}=-G\sqrt{-D_hD_p}\eqno(A5)
$$
where the single-particle energies are introduced as
$$
\varepsilon _p=e_p, \varepsilon _h=-e_h+G \eqno(A6)
$$
Thereafter the expressions for the phonon amplitudes are obtained
as
$$
X_p^\mu (2\varepsilon _p-E_\mu )=-\sum_{p^{\prime }}X_{p^{\prime
}}^\mu \widetilde{A}_{pp^{\prime }}+\sum_{h^{\prime }}Y_{h^{\prime
}}^\mu B_{ph^{\prime }}\Longrightarrow X_p^\mu
=\frac{-\sum_{p^{\prime }}X_{p^{\prime }}^\mu
\widetilde{A}_{pp^{\prime }}+\sum_{h^{\prime }}Y_{h^{\prime }}^\mu
B_{ph^{\prime }}}{2\varepsilon _p-E_\mu }\eqno(A7a)
$$
$$
Y_h^\mu (2\varepsilon _h+E_\mu )=-\sum_{p^{\prime }}X_{p^{\prime
}}^\mu B_{hp^{\prime }}+\sum_{h^{\prime }}Y_{h^{\prime }}^\mu
\widetilde{C}_{hh^{\prime }}\Longrightarrow Y_h^\mu
=\frac{-\sum_{p^{\prime }}X_{p^{\prime }}^\mu B_{hp^{\prime
}}+\sum_{h^{\prime }}Y_{h^{\prime }}^\mu \widetilde{C}_{hh^{\prime
}}}{2\varepsilon _h+E_\mu }\eqno(A7b)
$$
Due to the particle-hole symmetry, the removal mode satisfy
exactly the same equations. It means that
$$
X_p^{\mu}=X_{h=N-p+1}^{\lambda=\mu}, Y_h^{\mu}=Y_{p=N-h+1}^
{\lambda=\mu}\eqno(A8)
$$
Below we will use the notation $\mu$ for both modes.

Returning to the single particle occupation numbers $n_k$ let us
remind that in the picket fence model the following exact relation
between one body operator $N_k$ and two body operator $P_k^{+}P_k$
is verified for any non singly-occupied level $k$:
$$
N_k=2P_k^{+}P_k \eqno(A9)
$$
Taking the average of both parts of this relation  with respect to
the SCRPA vacuum state we obtain
$$
<N_p>=2D_p\sum_{\mu}(Y_p^{\mu})^2 \eqno(A10)
$$
On the other hand, using relations (A7) we can rewrite this
expression as
$$
<N_p>=2D_p\sum_{\mu}\frac{1}{\left( E_{\mu}+2\epsilon_p \right)^2}
\left[\sum_h
X_h^{\mu}B_{ph}-\sum_{p'}Y_{p'}^{\mu}\widetilde{C}_{pp'} \right]
\left[\sum_h
X_h^{\mu}B_{hp}-\sum_{p'}Y_{p'}^{\mu}\widetilde{C}_{p'p} \right]
\eqno(A11)
$$
It is easy to show that the fraction in this expression can be
expressed through the GFs as
$$
\frac{1}{\left( E_{\mu}+2\epsilon_p \right)^2}=-\lim_{t^{\prime
}-t\rightarrow 0^{+}} \int \int
 dt_1^{\prime }dt_1G_p^{0(t-t_1)}\theta (t_1^{\prime }-t_1)e^{-i(E_\mu
+\varepsilon _p)(t_1^{\prime }-t_1)}G_p^{0(t_1^{\prime }-t^{\prime
})} \eqno(A12)
$$
where
$$
G_p^{0(t-t_1)}=-i<T_tc_p(t)c_p^{+}(t_1)>=-i\theta
(t-t_1)e^{-i\varepsilon_p(t-t_1)} \eqno(A13)
$$
If we now take into account the spectral representation of the
pair operator $P_p$
$$
P_p(t)=D_p\left[ \sum_{\mu}e^{-iE_{\mu}t}Q_{\mu}X_p^{\mu}+
\sum_{\mu}e^{iE_{\mu}t}Q_{\mu}^{+}Y_p^{\mu} \right] \eqno(A14)
$$
and define the antichronological (time reversed) single particle
GF as
$$
G_{\overline{p}}^{0(t-t_1)}=-i<T_tc_{\overline{p}}^{+}(t)c_{\overline{p}}(t_1)>=
-i\theta (t_1-t)e^{i\varepsilon_p(t-t_1)} \eqno(A15)
$$
we can pass to the expression for the sp GF which gives occupation
numbers which are consistent within the frame of the SCRPA with
exact relation (A9)
$$
G_p^{\left( t -t^{\prime }\right) }=G_p^{0(t -t^{\prime })} +\int
\int dt_1dt_1^{\prime }G_p^{0(t
-t_1)}M^{SCRPA}_pG_p^{0(t_1^{\prime }-t^{\prime })}\eqno(A16)
$$
where $M^{SCRPA}_p$ is a single particle mass operator
$$
M^{SCRPA}_p= \sum_{k_1k_2}G_{\overline{p}}^{0(t_1^{\prime }-t_1)}\overline{{\cal H%
}}_{pk_1}^{(0)}G_{k_1k_2}^{(t_1^{\prime }-t_1)}\overline{{\cal H}}%
_{pk_2}^{(0)}\eqno(A17)
$$
and $\overline{{\cal H}}_{pk}^{(0)}$ is the renormalised effective
Hamiltonian which has the following form in terms of RPA matrixes
$A_{pp'}$ and $B_{ph}$
$$
\overline{{\cal
H}}_{pp'}^{(0)}=\sqrt{D_p}\widetilde{A}_{pp'}\frac{1}{\sqrt{D_{p'}}}
\eqno(A18a)
$$
$$
\overline{{\cal
H}}_{ph}^{(0)}=\sqrt{D_p}B_{ph}\frac{1}{\sqrt{-D_h}} \eqno(A18b)
$$
From (A16) and (A17) we then can define the occupation numbers as
usual.
 In general, the occupation numbers from (32) and
(A16), (A17) have slightly different values. This is due to the
fact that the ansatz (A1), (A2) for the RPA operators is too
restricted for a groundstate fulfilling $Q|0>=0$ to exist. It can,
however, be shown \cite{new} that the necessary corrections to
(A1), (A2) are small. Anyhow, the differences in occupation
numbers obtained from the two methods are a measure of the
importance of the terms neglected in (A1), (A2). The difference of
the mass operators (23) and (A17) is that in the latter both
vertices are dressed, whereas in (23) one vertex remains at the
unrenormalized value (G).

To find occupation numbers $<n_p>$ at finite temperature we adopt
the expressions obtained for single particle GF at zero
temperature (eq.(A16)-(A18)). To do it, it is necessary to change
all zero temperature GFs to the Matsubara ones and use for the
vertices the renormalised effective Hamiltonian (15)
$$
\overline{{\cal H}}_{pk}^{(0)}=\sqrt{D_p}\widetilde{{\cal
H}}_{pk}^{(0)} \frac{1}{\sqrt{\left|D_{k}\right|}} \eqno(A19)
$$
where
$$
\widetilde{\cal H}_{pk}^{(0)}={\cal
H}_{pk}^{(0)}-2\delta_{pk}\varepsilon_k
$$
and
$$
\varepsilon_k=e_k-Gf_k,
$$
$$
f_k=\frac{1}{1+e^{\varepsilon_k \beta}}
$$
Then we get for the single particle Matsubara GF
$$
G_p^{\left( \tau -\tau ^{\prime }\right) }=G_p^{0(\tau -\tau
^{\prime })}+ \int \int d\tau _1d\tau _1^{\prime }G_p^{0(\tau
-\tau _1)}M^{SCRPA}_p G_p^{0(\tau _1^{\prime }-\tau ^{\prime })}
$$
$$
=G_p^{0(\tau -\tau ^{\prime })}+\int \int d\tau _1d\tau _1^{\prime
}G_p^{0(\tau -\tau _1)}\left[
\sum_{k_1k_2}G_{\overline{p}}^{0(\tau _1^{\prime }-\tau _1)}\overline{{%
\mathcal{H}}}_{pk_1}^{(0)}G_{k_1k_2}^{(\tau _1^{\prime }-\tau _1)}\overline{{%
\mathcal{H}}}_{pk_2}^{(0)}\right] G_p^{0(\tau _1^{\prime }-\tau
^{\prime })} \eqno(A20)
$$
where
$$
G_p^{0(\tau -\tau ^{\prime })}=-e^{-\varepsilon _p(\tau -\tau
^{\prime })}\left[ \theta \left( \tau -\tau ^{\prime }\right)
(1-f_p)-\theta \left( \tau ^{\prime }-\tau \right) f_p\right] ,
$$
Taking this integral in the limit $\tau -\tau ^{\prime
}\longrightarrow 0^{-} $ we obtain the occupation numbers
 $n_p$ in the SCRPA at finite temperature
\[
n_{p}=<c_{p}^{+}c_{p}>=f_p+D_{p}\sum_{\mu }\left( Y_{p}^{\mu
}\right) ^{2}\left[ n_{B}\left( E_{\mu }\right) \left( 1-2f_p
\right) +\left( 1-f_p \right) ^{2}\right]
\]
\[
+D_{p}\sum_{\mu }\left( X_{p}^{\mu }\right) ^{2}\left[ n_{B}\left(
E_{\mu }\right) \left( 1-2f_p \right) -\left( f_p \right) ^{2}%
\right]
\]
\[
-f_p \left( 1-f_p \right) \beta D_{p}\left[ \sum_{\mu }\left(
Y_{p}^{\mu }\right) ^{2}\left( 1+n_{B}\left( E_{\mu }\right) -f_p
\right) \left( 2\varepsilon _{p}+E_{\mu }\right) \right]
\]
$$
-f_p \left( 1-f_p \right) \beta D_{p}\left[ \sum_{\mu }\left(
X_{p}^{\mu }\right) ^{2}\left( n_{B}\left( E_{\mu }\right) +f_p
\right) \left( 2\varepsilon _{p}-E_{\mu }\right) \right]
\eqno(A21)
$$

One should note here one thing about the direct use of the exact
relation (A9) for the definition of the $<N_p>$. In reality the
identity (A9) is only valid for the collective subspace (spanned
by the non singly-occupied levels) or, in other words, for levels
which have partial seniority $s_k=0$. But when we work in the
grand canonical ensemble, seniority of the level $s_k$ is not a
good quantum number, since averaging procedure over GCE mixes all
seniorities. This fact is reflected in the above expression (A21)
where we can see that level occupation numbers
$<N_p>=n_p+n_{\bar{p}}$ due to thermal factors $f_p$ (Fermi-Dirac
distribution) are not equal to $<2P^+_pP_p>$.

Before coming to a numerical example let us make a further
comment. Knowing the occupation numbers as a function of the
amplitudes $X$ and $Y$ as in (A21) or (32), a natural idea would
be to use this to express the ground state energy entirely as a
function of $X$ and $Y$. Then minimizing under the constraint of
the normalization conditions (21) leads to an equation determining
$X, Y$ amplitudes. Such a procedure has been proposed in the past
by Jolos and Rybarska \cite{Jol}. However, already in our example
where the occupation numbers are exactly known (at zero
temperature) as a function of $X, Y$ amplitudes, we have checked
that this leads to worse results than with the approach advanced
in the main text. In other examples the functional $n_p[X,Y]$ may
only be known approximately and then the minimization of the
ground state leads to deteriorated results, as we also have
checked. In the Green's function approach where one first
calculates excitation energies, i.e. energy differences, before
calculating the ground state energy, such uncertainties like the
precise knowledge of $n_p[X,Y]$ are minimized and the solution of
the SCRPA equations is therefore much more stable.

Let us now compare the results based on the two different
definitions of the occupation numbers (eq. (32) and (A21)). We
will denote results with eq. (32) by TSCRPA and results obtained
with the second definition (eq. (A21)) by TSCRPA1. All
calculations are made for the system with 10 particles on 10
levels. Results for the correlation energy (37) as a function of
the coupling constant $G$ are given in Tab. 1 and Tab. 2 for $T=0$
and $T=1$, respectively. In these tables we denote results
obtained with (A21) and (26) by TSCRPA1-1 and the ones obtained
with (A21) and (25) by TSCRPA1-2. We can see that TSCRPA
systematically gives underestimated results with respect to the
exact ones. Using definition (A21) we then can see that TSCRPA1-1
at zero temperature slightly overestimates the exact ground state
energy (this corresponds to the results given in \cite{DS1}) while
TSCRPA1-2 always underestimates it. At finite temperature all
approximations (TSCRPA and TSCRPA1) give close results which
underestimate the exact ones. In Tab. 3 and 4 we show excitation
energies of the first addition mode for $T=0$ and $T=1$. In both
cases, TSCRPA and TSCRPA1, excitation energies grow with
increasing $G$ for zero and nonzero temperatures reproducing quite
well the exact results. We can see that for the calculated
quantities (correlation and excitation energies) TSCRPA1-1 is
slightly closer to the exact results than TSCRPA.

\newpage


 Table 1. Correlation energy as function of $G$ at
zero temperature.

\vspace{1cm}

\begin{tabular}{|c|c|c|c|c|}
\hline
   G   & Exact  & TSCRPA1-1 & TSCRPA1-2 & TSCRPA \\
\hline
  0.10 & -0.036 & -0.037 & -0.036 & -0.036 \\
  0.20 & -0.167 & -0.169 & -0.153 & -0.159 \\
  0.30 & -0.435 & -0.445 & -0.342 & -0.379 \\
  0.33 & -0.551 & -0.564 & -0.408 & -0.461 \\
  0.34 & -0.593 & -0.608 & -0.430 & -0.489 \\
  0.35 & -0.638 & -0.654 & -0.453 & -0.518 \\
  0.36 & -0.685 & -0.702 & -0.476 & -0.548 \\
  0.40 & -0.896 & -0.917 & -0.569 & -0.670 \\
\hline
\end{tabular}

\vspace{1cm}
Table 2. Correlation energy as function of $G$ at $T=1$.

\vspace{1cm}

\begin{tabular}{|c|c|c|c|c|}
\hline
   G   & Exact  & TSCRPA1-1 & TSCRPA1-2 & TSCRPA \\
\hline
  0.10 & -0.030 & -0.029 & -0.018 & -0.029 \\
  0.20 & -0.142 & -0.126 & -0.082 & -0.122 \\
  0.30 & -0.372 & -0.299 & -0.208 & -0.295 \\
  0.33 & -0.476 & -0.367 & -0.259 & -0.365 \\
  0.34 & -0.515 & -0.392 & -0.278 & -0.391 \\
  0.35 & -0.556 & -0.418 & -0.297 & -0.417 \\
  0.36 & -0.600 & -0.444 & -0.317 & -0.445 \\
  0.40 & -0.803 & -0.559 & -0.403 & -0.569 \\ \hline
\end{tabular}

\vspace{1cm}

Table 3. Excitation energy of the first addition mode as function
of $G$ at $T=0$.

\vspace{1cm}

\begin{tabular}{|c|c|c|c|}
  \hline
   G   & Exact  & TSCRPA1 & TSCRPA \\
\hline
  0.10 & 1.001 & 1.001 & 1.001 \\
  0.20 & 1.005 & 1.012 & 1.012 \\
  0.30 & 1.014 & 1.049 & 1.053 \\
  0.33 & 1.018 & 1.068 & 1.074 \\
  0.34 & 1.019 & 1.075 & 1.081 \\
  0.35 & 1.022 & 1.082 & 1.089 \\
  0.36 & 1.023 & 1.089 & 1.098 \\
  0.40 & 1.031 & 1.123 & 1.136 \\ \hline
\end{tabular}

\vspace{1cm}
\newpage
Table 4. Excitation energy of the first addition mode as function
of $G$ at $T=1.0$.

\vspace{1cm}

\begin{tabular}{|c|c|c|c|}
\hline
   G   & TSCRPA1 & TSCRPA \\
\hline
  0.10 & 1.017 & 1.030 \\
  0.20 & 1.075 & 1.126 \\
  0.30 & 1.175 & 1.281 \\
  0.40 & 1.299 & 1.476 \\
  0.41 & 1.312 & 1.497 \\
  0.42 & 1.324 & 1.518 \\
  0.43 & 1.337 & 1.539 \\
  0.44 & 1.349 & 1.560 \\
  0.45 & 1.362 & 1.581 \\
  \hline
\end{tabular}

\newpage
\begin{figure}[tph]
\centering
\includegraphics[scale=0.6]{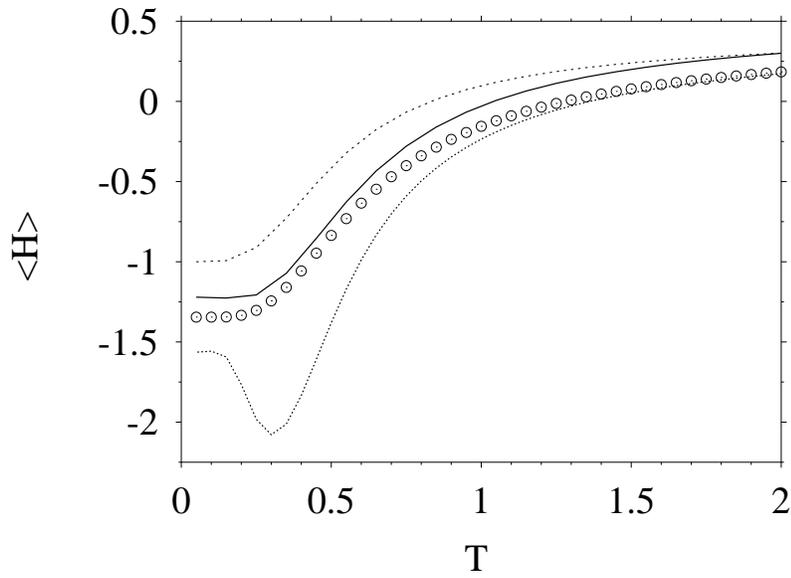}
\vspace{0.5cm} \caption{ The average energy $<H>$ as a function of
the temperature for $\Omega=N=2$ and $G=0.9$. The exact results --
open circles; the TMFA results -- dotted line; the TRPA results --
dashed line and the TSCRPA results -- solid line. }
\end{figure}

\begin{figure}[tph]
\centering
\includegraphics[scale=0.6]{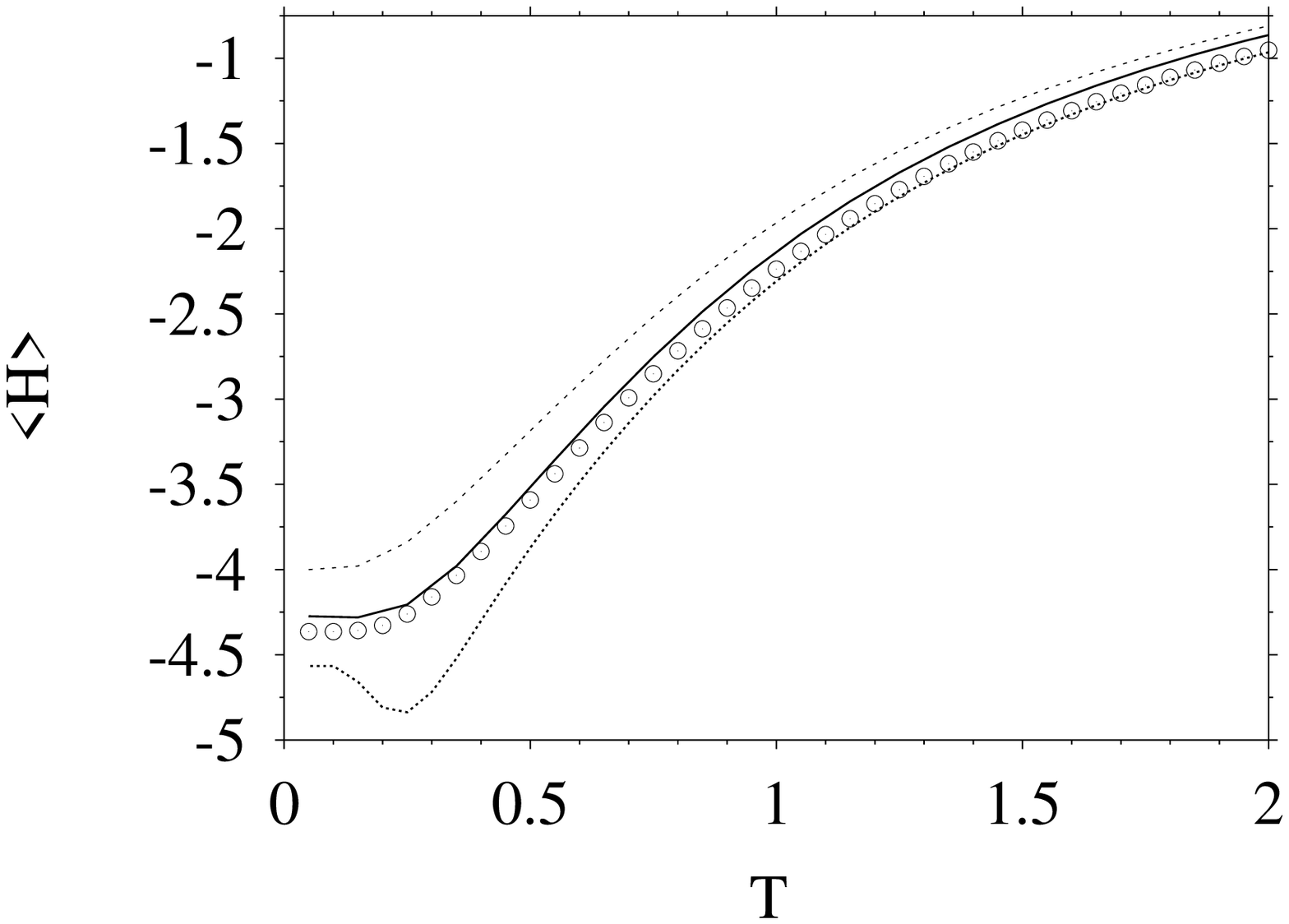}
\vspace{0.5cm} \caption{ The average energy $<H>$ as a function of
the temperature for $\Omega=N=4$ and $G=0.5$. For notation, see
Fig.2 }
\end{figure}

\begin{figure}[tph]
\centering
\includegraphics[scale=0.6]{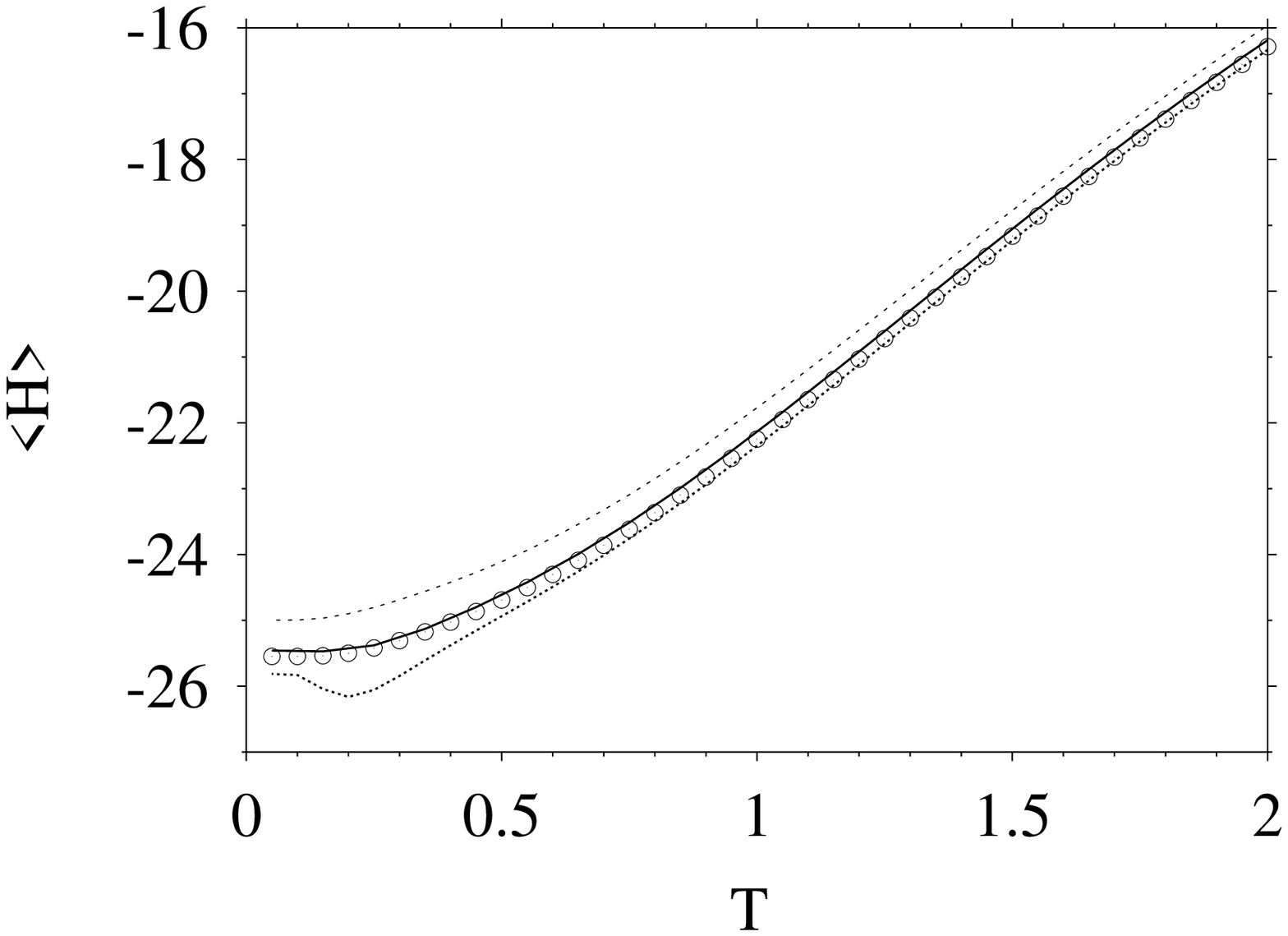}
\vspace{0.5cm} \caption{ The average energy $<H>$ as a function of
the temperature for $\Omega=N=10$ and $G=0.33$. For notation, see
Fig.2.}
\end{figure}

\begin{figure}[tph]
\centering
\includegraphics[scale=0.6]{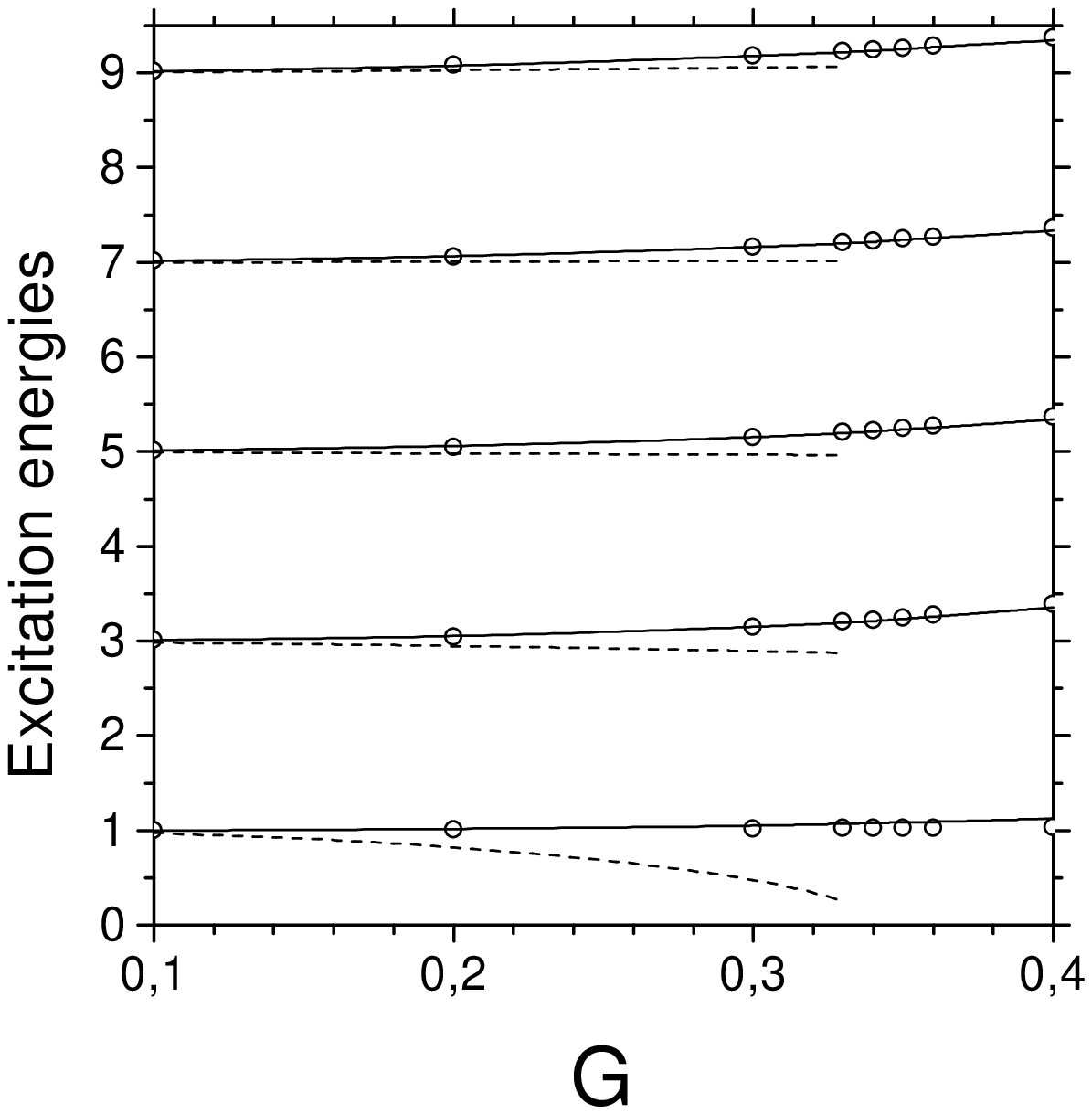}
\vspace{0.5cm} \caption{ Excitation energies as a function of the
interaction constant $G$ for $\Omega=10$ and $T=0$. Notations: the
exact results -- open circles; TRPA results -- dashed lines and
the TSCRPA results -- solid lines. }
\end{figure}

\begin{figure}[tph]
\centering
\includegraphics[scale=0.6]{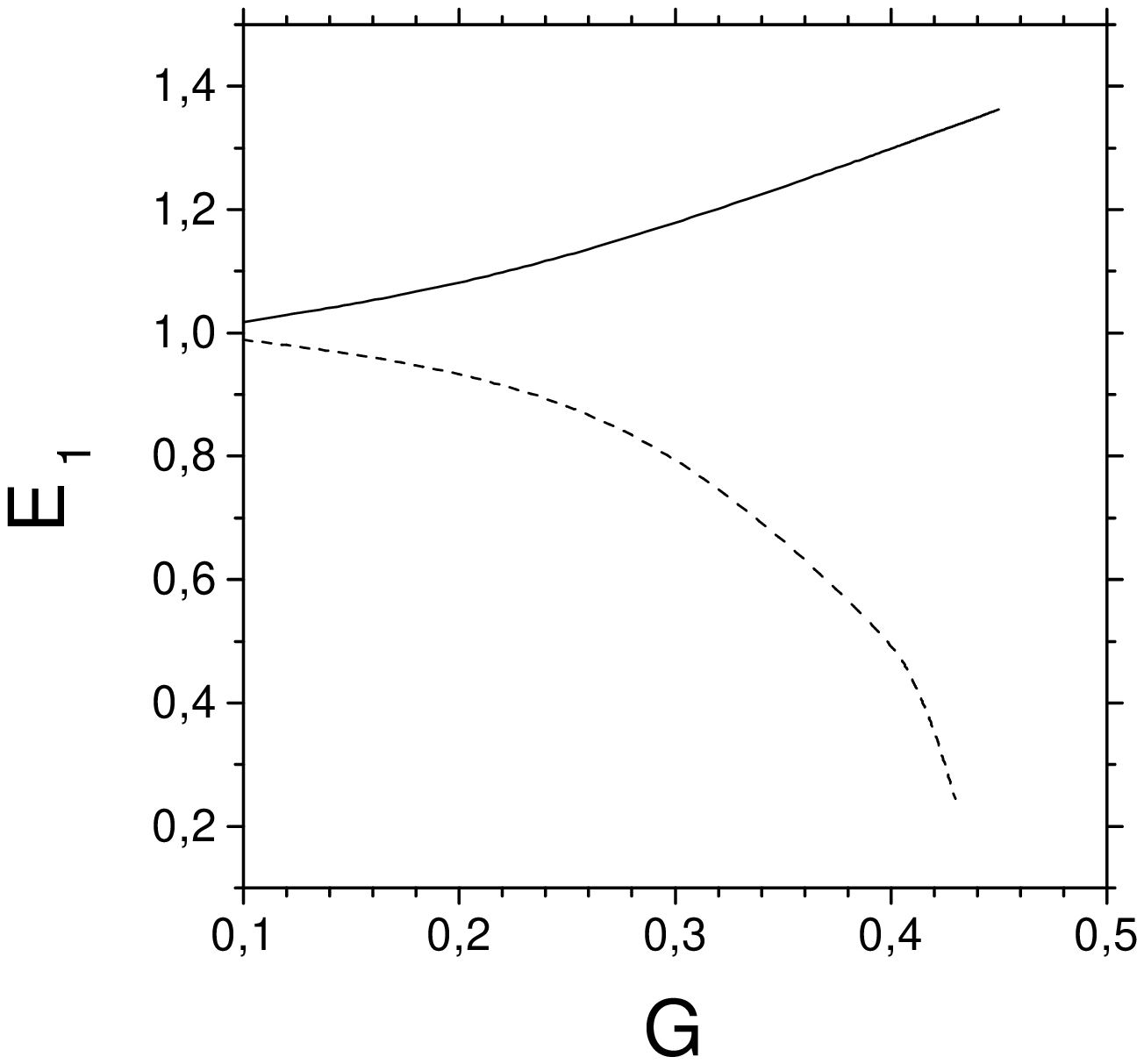}
\vspace{0.5cm} \caption{ The first excitation energy $E_1$ as a
function of the interaction constant $G$ for $\Omega=10$ and
$T=0.5$. Notations: TRPA results -- dashed line, the TSCRPA
results -- solid line. }
\end{figure}

\begin{figure}[tph]
\centering
\includegraphics[scale=0.6]{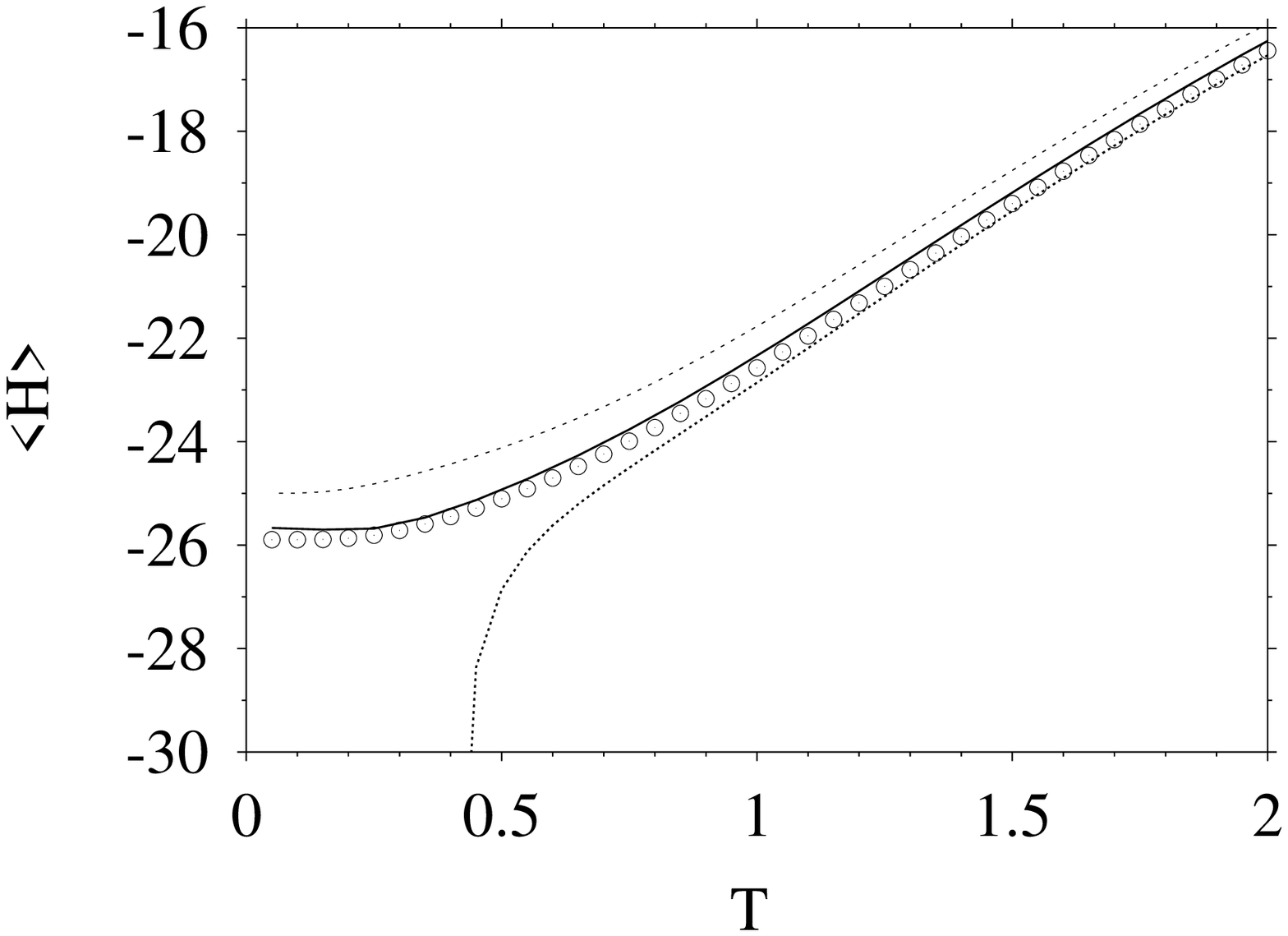}
\vspace{0.5cm} \caption{ The average energy $<H>$ as a function of
the temperature for $\Omega=N=10$ and $G=0.4$. For notation, see
Fig.2.}
\end{figure}

\begin{figure}[tph]
\centering
\includegraphics[scale=0.6]{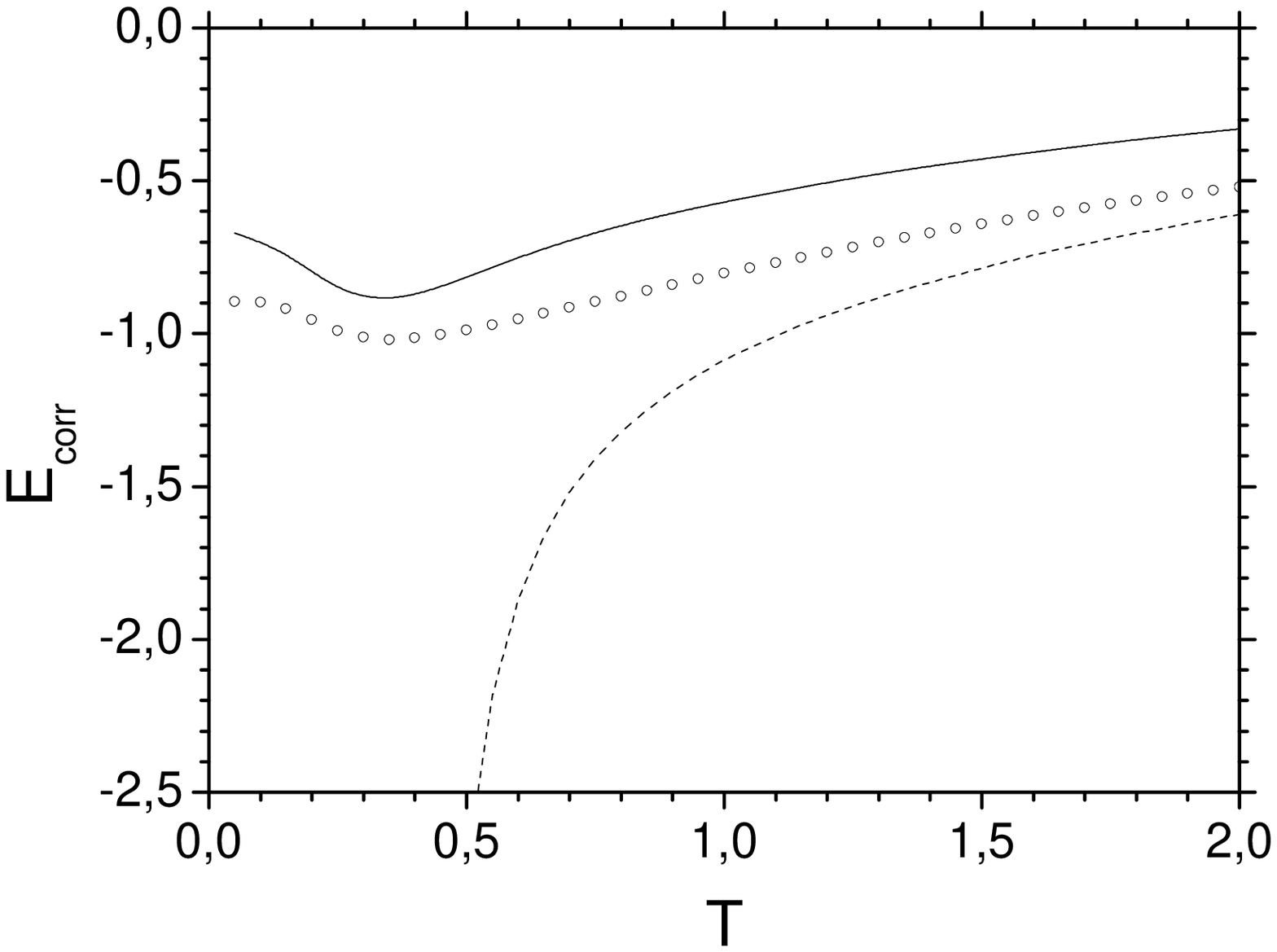}
\vspace{0.5cm} \caption{ The correlation energy $E_{corr}$ as a
function of the temperature for $\Omega=N=10$ and $G=0.4$. For
notation, see Fig.2. }
\end{figure}

\begin{figure}[tph]
\centering
\includegraphics[scale=0.6]{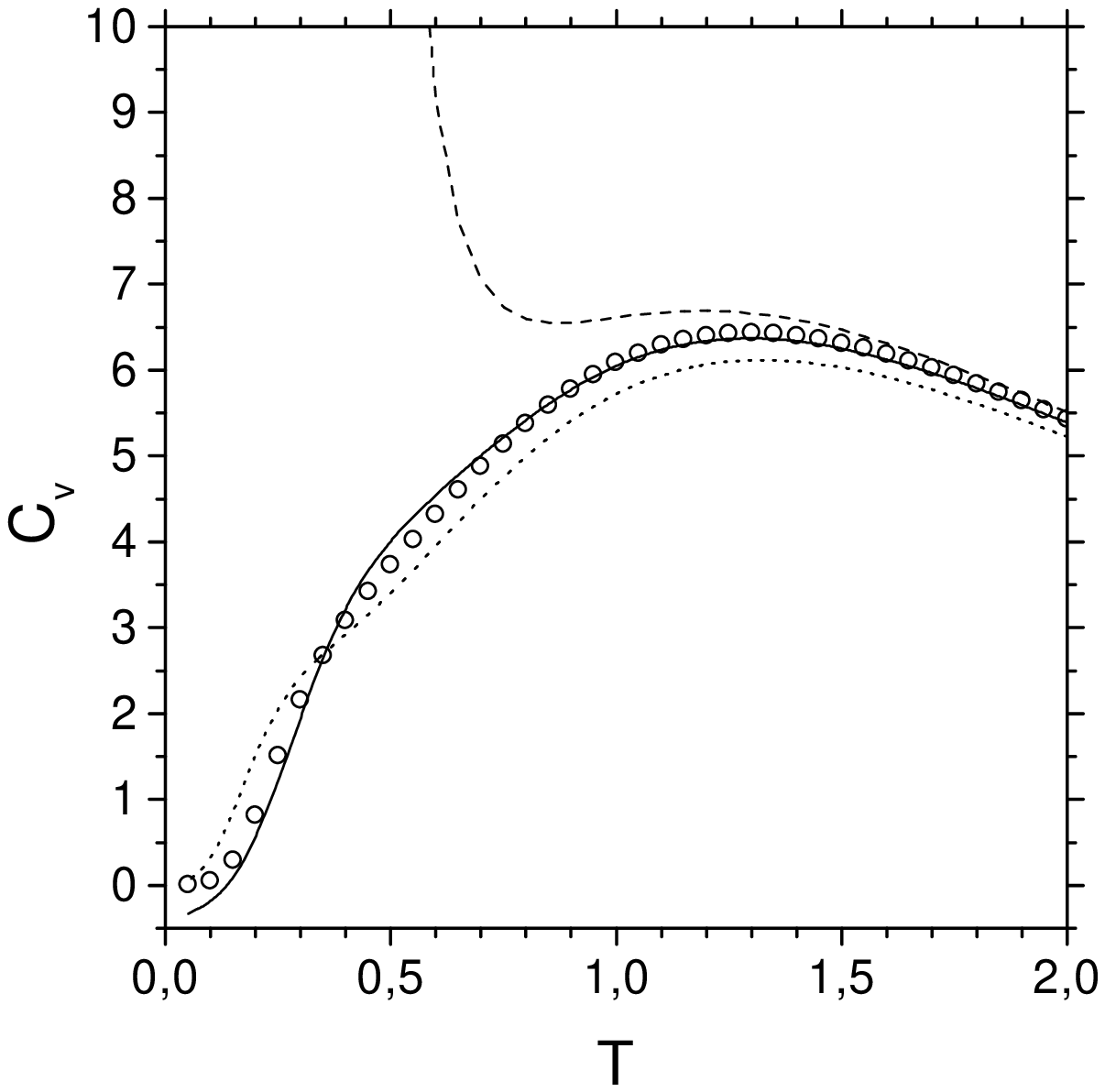}
\vspace{0.5cm} \caption{ The heat capacity $C$ as a function of
the temperature for $\Omega=N=10$ and $G=0.4$. For notation, see
Fig.2. }
\end{figure}

\begin{figure}[tph]
\centering
\includegraphics[scale=1.0]{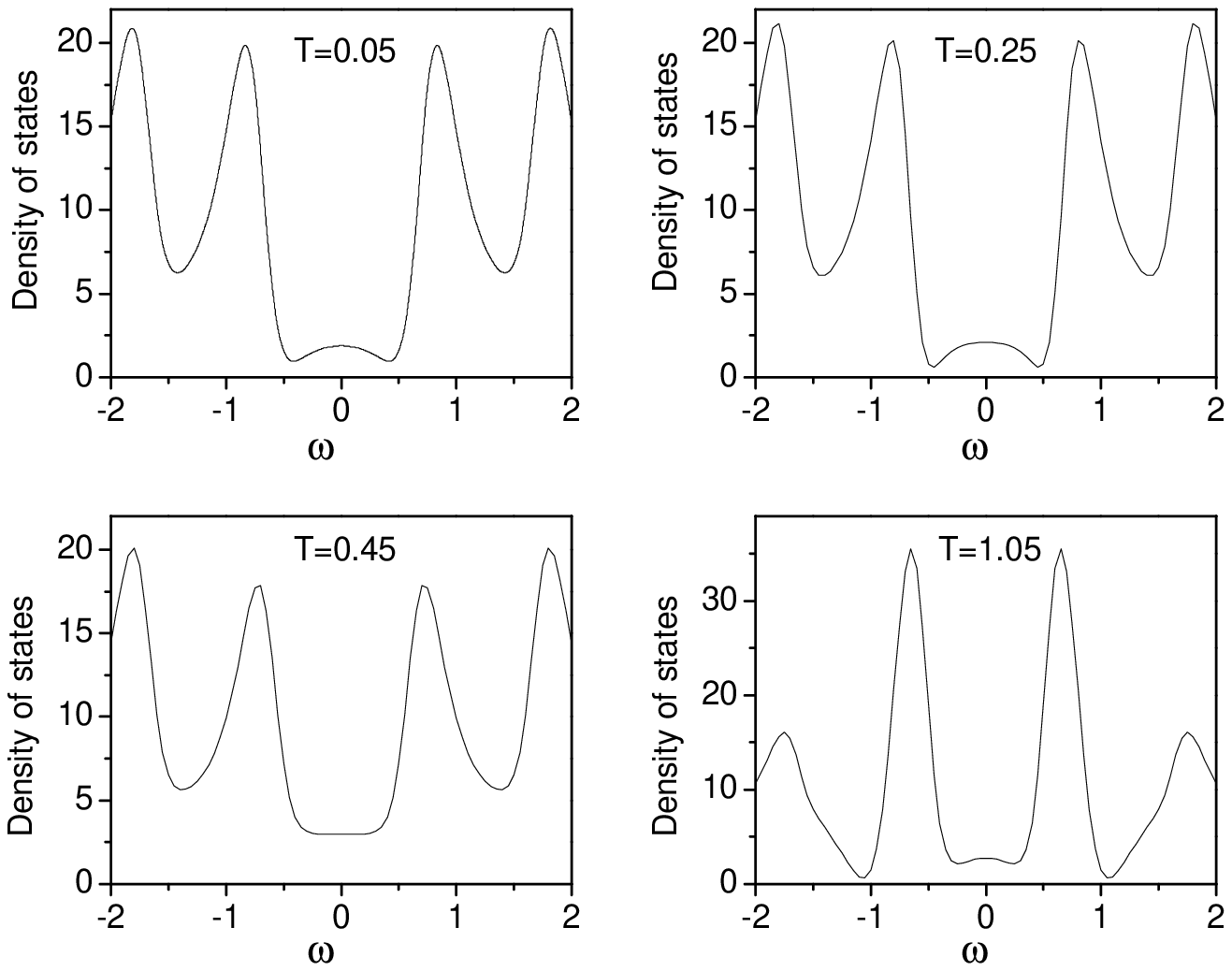}
\vspace{0.5cm} \caption{ The density of states as a function of
the frequency $\omega$ for $\Omega=N=10$, $G=0.4$ and $T=0.05$,
$T=0.25$, $T=0.45$ and $T=1.05$. }
\end{figure}

\begin{figure}[tph]
\centering
\includegraphics[scale=0.5]{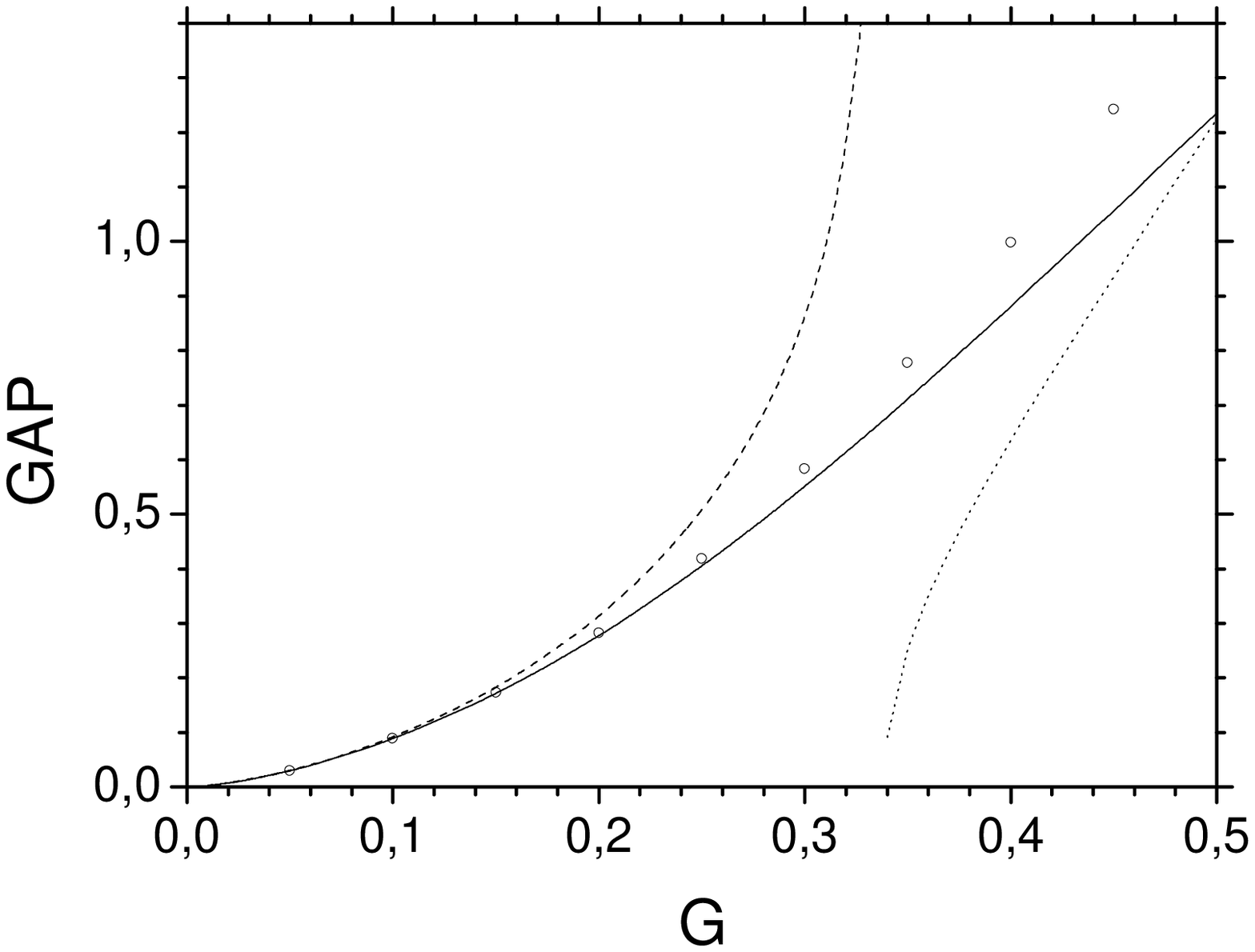}
\vspace{0.5cm} \caption{ The effective gap $\Delta$ as a function
of the interaction constant $G$ for $\Omega=N=10$ and $T=0$. The
exact results -- open circles; the BCS results -- dotted line; the
RPA results -- dashed line and the SCRPA results -- solid line. }
\end{figure}

\begin{figure}[tph]
\centering
\includegraphics[scale=0.6]{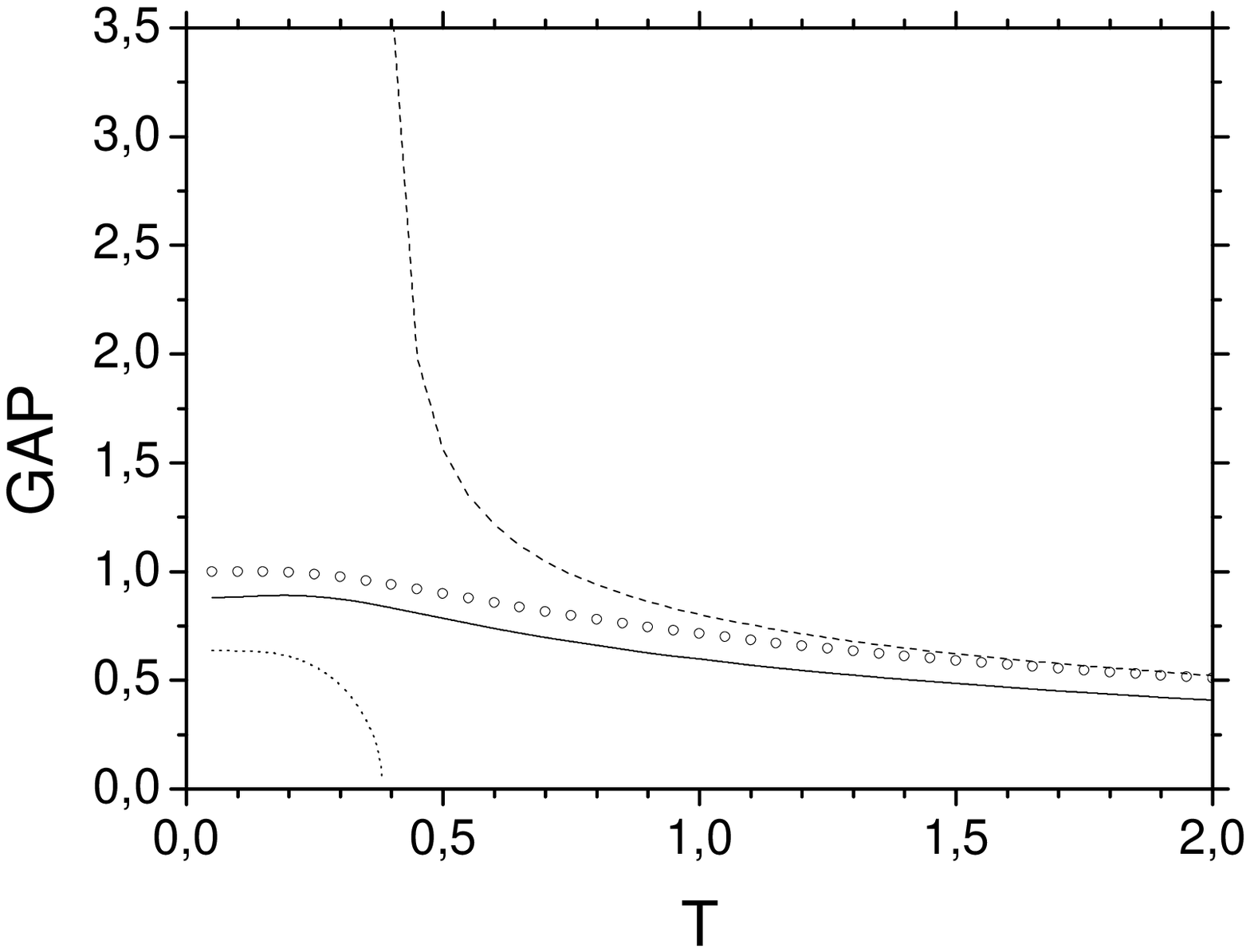}
\vspace{0.5cm} \caption{ The effective gap $\Delta$ as a function
of the temperature $T$ calculated for $\Omega=N=10$ and $G=0.4$.
For notation, see Fig.11. }
\end{figure}

\begin{figure}[tph]
\centering
\includegraphics[scale=0.5]{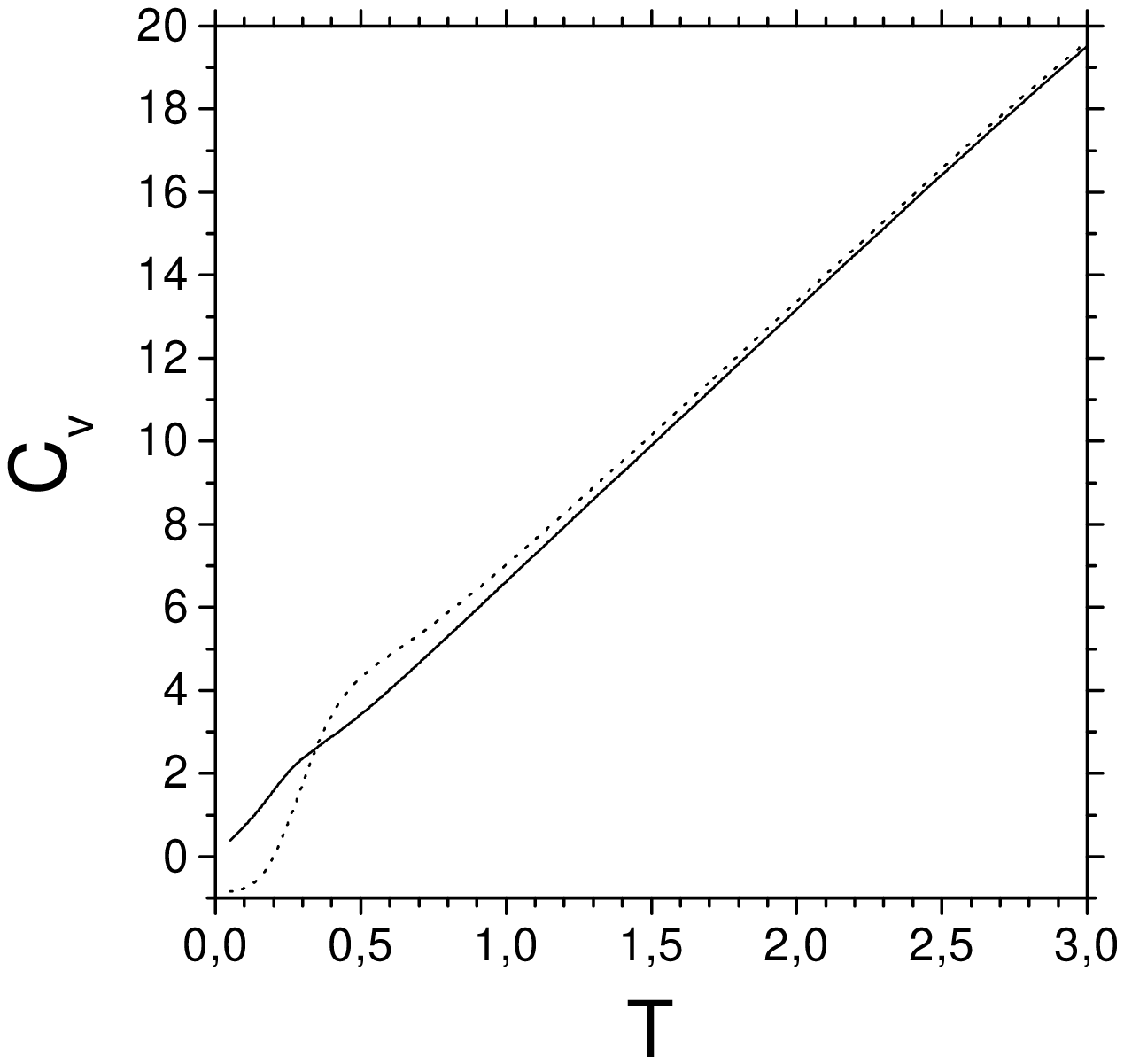}
\vspace{0.5cm} \caption{ The specific heat $C_v$ as a function of
the temperature calculated in the thermal SCRPA for $\Omega=N=50$.
Solid line corresponds to $G=0.128$ ($\delta/\Delta=50$) and
dotted line corresponds to $G=0.256$ ($\delta/\Delta=1$). }
\end{figure}

\end{document}